\documentclass[journal]{IEEEtran}
\usepackage{graphicx}
\usepackage[cmex10]{amsmath}
\usepackage{array}
\usepackage[hyphens]{url}
\usepackage{multirow}
\usepackage{array}
\usepackage{makecell}
\usepackage{threeparttable, tablefootnote}
\usepackage{amsmath,amsfonts,amssymb}
\usepackage{mathtools}
\usepackage{xcolor}
\DeclarePairedDelimiter{\ceil}{\lceil}{\rceil}
\DeclarePairedDelimiter\floor{\lfloor}{\rfloor}
\newcolumntype{P}[1]{>{\centering\arraybackslash}p{#1}}
\newcolumntype{M}[1]{>{\centering\arraybackslash}m{#1}}
\usepackage{algorithm}
\usepackage[noend]{algpseudocode}
\makeatletter
\def\BState{\State\hskip-\ALG@thistlm}
\makeatother

\usepackage{cite}
\ifCLASSINFOpdf
\else
\fi
\begin{document}
\title{A $51.3$ TOPS/W, $134.4$ GOPS In-memory Binary Image Filtering in $65$nm CMOS}

\author{Sumon Kumar Bose,~\IEEEmembership{Graduate~Student~Member,~IEEE,}

Deepak Singla,~\IEEEmembership{Member,~IEEE,}
 and Arindam Basu,~\IEEEmembership{Senior~Member,~IEEE}

\thanks{S. K. Bose is with Quasistatics inc., USA. He was with the School of EEE, NTU, Singapore 639798.  (e-mail: bose0003@e.ntu.edu.sg). This work
was done during his time at NTU.}
\thanks{A.Basu is with department of EE, City University of Hong Kong. He was earlier with NTU, Singapore. (e-mail: arinbasu@cityu.edu.hk).}
\thanks{D. Singla is with department of Bioengineering, university of UCLA. This work
was done during his time at NTU.}
\vspace{-2em}
%\thanks{Z. Lei is with the Centre of Excellence %in IC
%Design (VIRTUS), School of EEE, NTU, Singapore 639798 (e-mail: CharlesZhang@ntu.edu.sg).}

%\thanks{A. Patil is associated with Tork Motors, Pune, India 411026.}% <-this % stops a space
}
\maketitle
\begin{abstract}
%Image denoising is one of the utmost image processing steps in the application of traffic surveillance, object detection, and tracking in the scene. 
%Image denoising is the first step in the processing pipeline for most video applications. Execution of such processes on edge devices demands low latency due to the live video streaming, and low energy consumption to prolong battery life. 
Neuromorphic vision sensors (NVS) can enable energy savings due to their event-driven  that exploits the temporal redundancy in video streams from a stationary camera. However, noise-driven events lead to the false triggering of the object recognition processor. Image denoise operations require memory-intensive processing leading to a bottleneck in energy and latency. In this paper, we present in-memory filtering (IMF), a $6$T-SRAM in-memory computing based image denoising for event-based binary image (EBBI) frame from an NVS. We propose a non-overlap median filter (NOMF) for image denoising. An in-memory computing framework enables hardware implementation of NOMF leveraging the inherent read disturb phenomenon of $6$T-SRAM. To demonstrate the energy-saving and effectiveness of the algorithm, we fabricated the proposed architecture in a $65$nm CMOS process. As compared to fully digital implementation, IMF enables $>70\times$  energy savings and a $>3\times$ improvement of processing time when tested with the video recordings from a DAVIS sensor and achieves a peak throughput of $134.4$ GOPS. Furthermore, the peak energy efficiencies of the NOMF is $51.3$ TOPS/W, comparable with state of the art in-memory processors. We also show that the accuracy of the images obtained by NOMF provide comparable accuracy in tracking and classification applications when compared with images obtained by conventional median filtering.
\end{abstract}

% Note that keywords are not normally used for peer review papers.
\begin{IEEEkeywords}
In-memory computing, neuromorphic vision sensors, median filter, image denoising, address event representation.
\end{IEEEkeywords}

% \fi
\IEEEpeerreviewmaketitle

\section{Introduction}

%\IEEEPARstart{I}{n} recent years, we have witnessed a rapid proliferation of the internet of things (IoT) accelerated by the emerging $5$G  communication technology. Among all the sensors, the camera will contribute $95\%$ of the total internet traffic~\cite{IOVT} by $2030$. Consequently, the processing of data from the camera provides unique challenges and opportunities to the researchers. Limited bandwidth and battery capacity of the sensors node prevent frequent data transmission to the cloud. To address these challenges, edge computing frameworks~\cite{ADEPOS},~\cite{edge} have been proposed in the literature resulting in more computing energy consumption on edge devices. Hence, a paradigm shift is necessary for the architecture of sensors and sense-makers to reduce the sensing as well as the processing energy. 

\IEEEPARstart{T}{he} proliferation of sensors in the Internet of Things (IoT) combined with emergence of ultra low-power electronics and almost pervasive connectivity has marked the emergence of many new applications. Image-sensors producing video streams hold an unique position among all the sensors due to both the rich information carried in it as well as the requirement of extremely high network bandwidth and concomitant energy needed for its transmission~\cite{IOVT_jrnl}. Edge computing~\cite{edge} offers an alternative by processing the information locally using advanced machine learning techniques like deep neural networks (DNN); however, most state of the art DNN models are memory and compute intensive requiring significant energy dissipation as well~\cite{yolo}.  
\begin{figure}[t]
\centering
\includegraphics[width=0.45\textwidth]{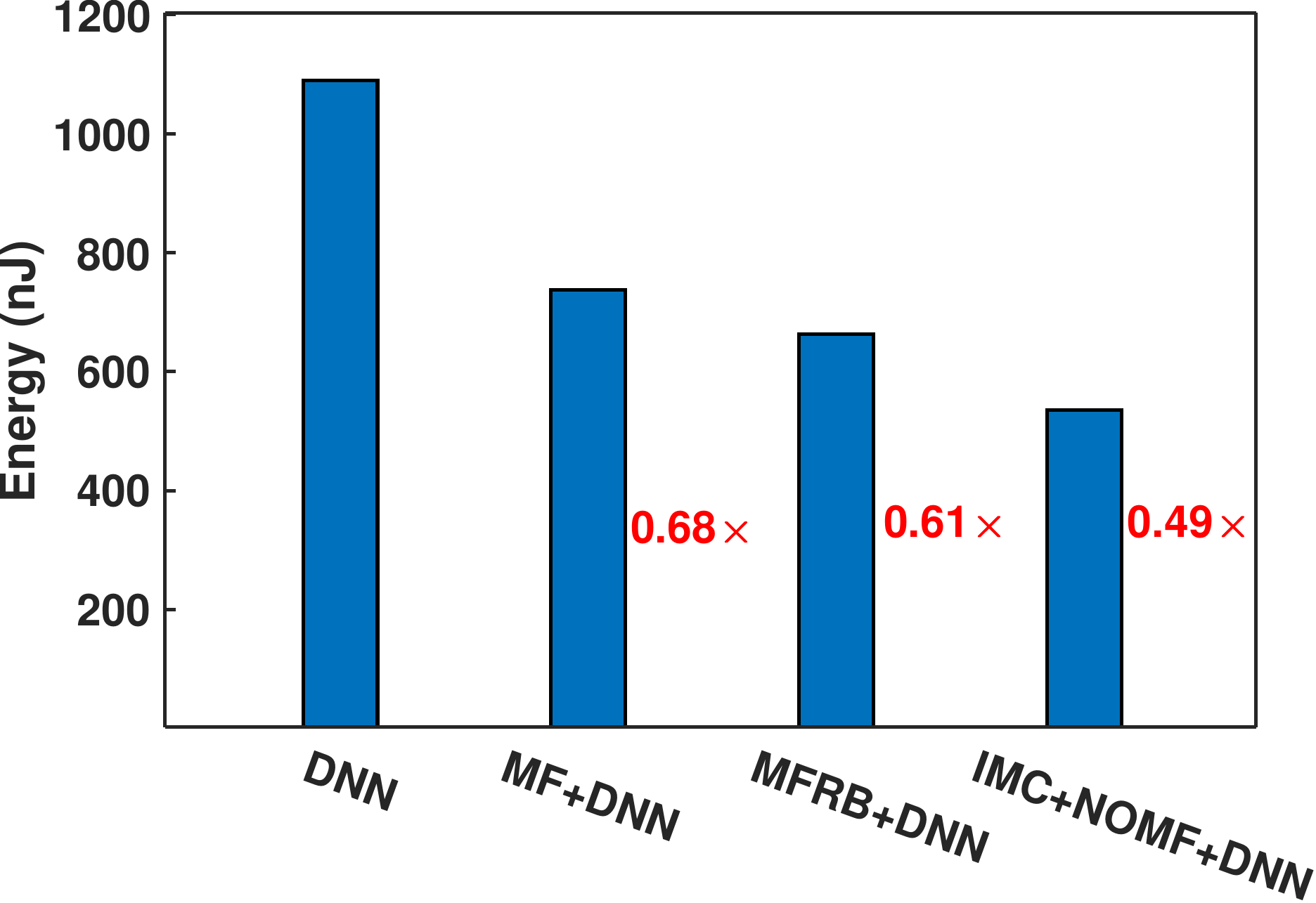}
\vspace{-1.0em}
\caption{
The energy required to run a DNN for object recognition on each frame can be reduced by not operating on empty frames which comprises $\approx 51\%$ of all frames. Including a conventional median filter to denoise the images requires significant energy overhead while IMC based approximate median filter enables exploiting the temporal redundancy fully.
%Essential steps of an object tracking system. (a) Read image frame: sample NVS memory at $t_f$ interval and write the EBBI frame to the SRAM/DRAM of a processor, (b) image denoising: removal of background noise, (c) region proposal: finding out the lower ($x_l$,$y_l$) and higher ($x_h$,$y_h$) coordinates of the bounding box (bbox) encapsulating the object, (d) object classification: classifying the objects specified in the bboxes, and (e) object tracking: tracking the object specified in the bbox across multiple frames and assigning a tracking id if a new object appears in the frame.
}
%\label{figSteps}
\label{fig:energy_denoise}
\vspace{-1.5em}
\end{figure}

One option to reduce computations is to downsample the image--however, this leads to loss in accuracy. Another option for stationary cameras is to exploit the temporal redundancy in video streams, i.e. background information is not changing from frame to frame. Bio-inspired neuromorphic vision sensors (NVS)~\cite{Posch2014},~\cite{Berner2013} that detect change in temporal contrast at the pixel level are ideally suited for this purpose since the pixels generate events only on detecting changes~\cite{aer}. However, in real scenarios, many spurious events are generated due to noise which requires the use of noise filtering or denoise operations~\cite{nnfilt},~\cite{Padala2018}. While event-driven filtering~\cite{nnfilt},~\cite{Padala2018} was proposed initially, hybrid frame-event approach~\cite{Jyotibdha_EBBIOT} is more suited for IoT operations. Since the event generated in the NVS will not reset until it is readout, they propose to sample the sensor memory in a burst after a regular interval $t_f$, to create an event-based binary image (EBBI) frame.  It allows the processor to reduce energy consumption by duty cycling once the processing ends. This also simplifies the noise filtering to a median filter operation and frames without valid objects (with all pixel values equal to $0$ due to absence of events) need not be processed by the object recognition DNN. 

Fig. \ref{fig:energy_denoise} shows the potential energy benefits from this approach. The classification energy of a DNN processor is assumed to be $1076.6$nJ based on the average of numbers reported in recent works~\cite{8226999,8588363,nverma} in the range of $730$nJ to $1700$nJ. Analyzing the frames from the traffic dataset in~\cite{singla2020ebbinnot}, we find $\approx 51\%$ of frames to be empty (frames without a valid object). Using energy estimates from measurements of the designed $320\times240$ memory read ($0.916$pJ/bit) and write energy ($6$pJ/bit) for a conventional digital implementation of median filter, we find that the average energy requirement reduces by only $32-39\%$ due to the high energy requirement of the median filter itself which in turn is largely due to memory access. In contrast, emerging in and near-memory computing framework eliminates the data transfer cost pointing to potential for great improvement.

In near or in-memory computing (IMC), the processing is done either on the boundary~\cite{Sylvester} or inside the memory. It not only gets rid of the data movement but also enables highly
parallel processing due to its simultaneous access to multiple cells~\cite{Biswas2018_thesis}. In general, the SRAM-based realization of near or in-memory computing  can be categorized as charge sharing~\cite{ABIS},~\cite{Murmann}, current summation~\cite{Verma},~\cite{twt}, delay-based~\cite{sandwitch} and xor-based digital~\cite{Kim} implementation. While the majority of the works on IMC are focused on the classification of the objects, in this paper, we use SRAM-based IMC for efficient denoising of the event-based binary image (EBBI). Fig. \ref{fig:energy_denoise} shows that our approach allows full exploitation of the temporal redundancy in video data enabling average energy per frame to go down by $\approx 51\%$ equivalent to the sparsity in input data.

In this paper, we propose a non-overlap median filter (NOMF) for image denoising. Even though the NOMF introduces approximation in computation~\cite{ADEPOS}, it: 1) reduces the number of computes, explained later in section~\ref{performance}, and 2) enables hardware implementation of image denoising, leveraging inherent read disturb phenomenon of $6$T-SRAM.
Our significant contributions in this paper are as follows: \begin{enumerate}
\item We propose a $6$T-SRAM in-memory computing-based image denoising integrated circuit for event-based binary images leveraging read disturb of SRAM. This article is an extended version of~\cite{bose202075kb}, providing the characterization and measurement results of NOMF in four fabricated chips. The proposed NOMF enables $\sim114\times$ energy savings compared to a digital counterpart when tested with the video recordings from a DAVIS sensor~\cite{davis_camera} and achieves a peak throughput of $0.58$ frames/$\mu$s for $240\times180$ binary images.
\item We evaluate the performance of an overlap-based tracker as well as an object classifier (LeNet inspired Convolutional neural network) using images from the median filter and NOMF. The performance of the object classifier and overlap-based tracker for both images is comparable. %Only $2.6\%$ degradation in the area under the curve of F1-score plot is observed due to the approximation in the NOMF approach.
\item We characterize the unwanted bit flip scenario and discuss different design aspects to prevent it in detail.
\item Also, a near memory computing circuit is proposed for valid frame detection tracking the bit line voltages during the filtering operation.
\end{enumerate}

We organize the remaining part of the paper as follows. In section~\ref{filter}, we review the basics of a median filter and proposed non-overlapped median filter. We discuss the IMF processor architecture and present a $6$T SRAM-based hardware implementation of the non-overlap median filter in section \ref{overall_archi}. Section \ref{hwresults} presents measurement results and performance evaluation and comparison of the proposed NOMF with the conventional median filter in the applications of object tracking and detection. In section \ref{validframedetection}, we propose a near memory computing circuit for valid frame detection, followed by a conclusion in section \ref{conclu}. 

\section{Image Denoising}
\label{filter}
In this section, we will give a brief overview of a median filter and the proposed approach of noise filtering.
\subsection{Median Filter}
A median filter slides an $n\times n$ window over an input image and replaces the center pixel of the $n\times n$ output patch by the median value of the $n^2$ pixels associated with the window~\cite{mfilter}. We can express the nonlinear operation of the median filter mathematically as Eq. (\ref{eqn1}) where i, j $\subset \mathbb{Z}^+$ and $P_{mf}(i,j)$ denotes the filtered pixel at (i, j) position.
  \begin{equation}
 \begin{split}
P_{mf}(i,j) = &median\big(\big\{P(i+k,j+l) ~|~k,l \subset \mathbb{Z} \\& \text{and}~\in \big[-\frac{n-1}{2},\cdots,\frac{n-1}{2}\big]\big\}\big)
\label{eqn1}
\end{split}
\end{equation}

Usually, a realization  of the median filter requires sorting an array carrying $n^2$ pixels up to the mid-position. However, for a binary image, the implementation is quite simple and involves counting the number of occurrences of ``1" in an $n\times n$ patch. If the count equals or exceeds $\lceil \frac{n^2}{2} \rceil$, a ``1" is assigned to the center pixel of the filtered image. The mathematical operation of a median filter for a binary image follows Eq. (\ref{eqn2anew})  
\vspace{-0.2em}
\begin{align}
 P_{mf}(i,j)&=
\begin{dcases}
    1,& \text{if $\Sigma P(i+k,j+l) \geq \ceil*{\frac{n^2}{2}}$ }\\
    0,& \text{otherwise} \label{eqn2anew}
\end{dcases}
\end{align}
where $\ceil*{.}$ rounds a number to its nearest higher integer.

\begin{figure}[t]
\centering
\includegraphics[scale=0.50]{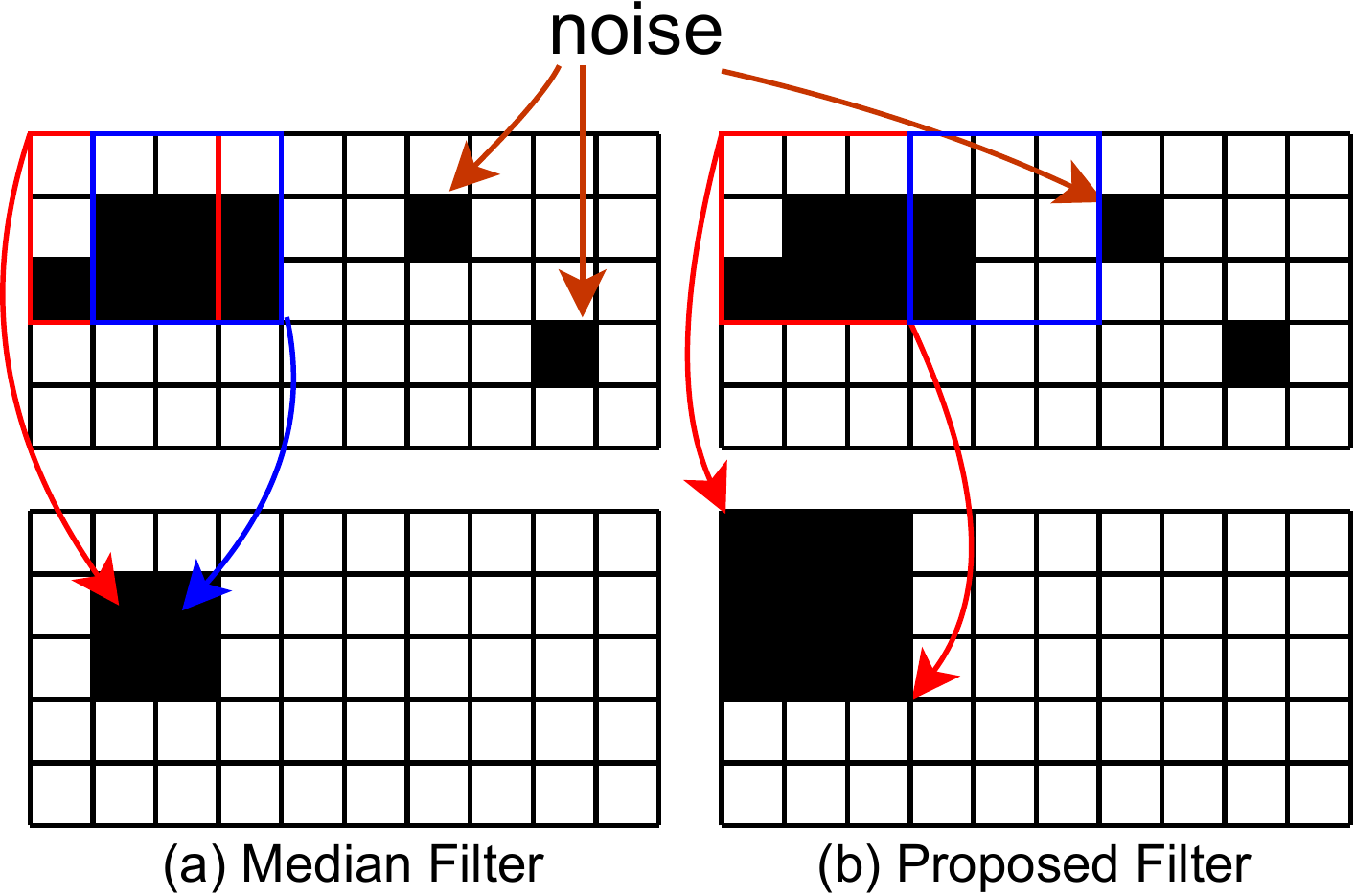}
\vspace{-0.5em}
\caption{The black dot represents a non-zero pixel of an image frame with binary values. (a) Conventional median filter using a $3\times3$ window and stride = $1$ (b) Proposed NOMF using a $3\times3$ window and stride = $3$ for image denoising. In NOMF, decisions for all $9$ pixels are taken simultaneously.}
\label{nonoverlap}
\vspace{-1.5em}
\end{figure}

\subsection{Proposed Non-overlap Median Filter}
In a conventional median filter, an $n\times n$ window slides over an image in an overlap fashion where the stride, $s=1$ as shown in Fig.~\ref{nonoverlap}(a). A simple implementation of the same following von Neumann architecture requires $n^2$+1 clock cycles and associated energy to fetch and sum up $n^2$ pixels bit by bit followed by  a comparison and a write operation in a separate memory. In contrast, since the neighboring pixels of an image have similar characteristics, we propose to apply the decision of an $n\times n$ window to all the $n^2$ pixels instead of the center one. This is equivalent to having stride $s=n$ (Fig. \ref{nonoverlap}(b)), resulting in a non-overlap median filter (NOMF) that we use in this work. While the proposed approach changes the object boundary slightly, it reduces the processing and memory access time by a factor of $n^2$ and enables the same memory to be utilized to store the filtered image. Furthermore, it facilitates circuit-level filter implementation for image denoising leveraging read disturb phenomenon of a $6$T-SRAM.

\section{In-memory Filtering (IMF)}
\label{overall_archi}
This section describes the top-level architecture of in-memory filtering, and the detailed circuit-level implementation of the proposed NOMF utilizing the read disturb phenomenon of a $6$T-SRAM. Performance analysis of the proposed method is presented at the end of this section.
\subsection{System Overview}
\begin{figure}[t]
\centering
\includegraphics[scale=0.5]{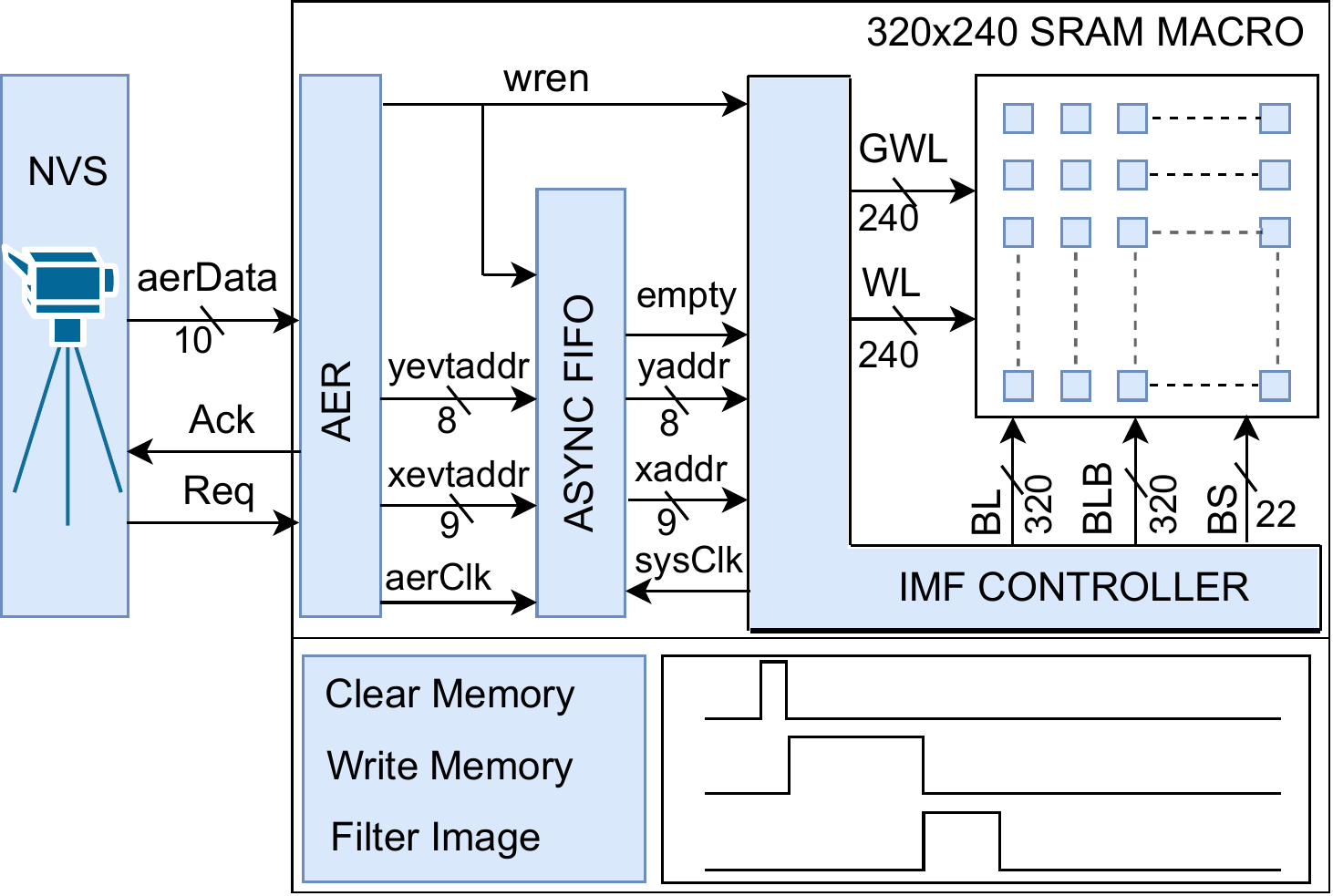}
\vspace{-0.4em}
\caption{Top-level architecture of the IMF processor. The chip consists of an address event representation module (AER), a $128\times32$ bit asynchronous buffer, an IMF controller, and a $320\times240$ SRAM macro. The waveform at the bottom of the image shows the different operations sequentially performed by the IMF processor.}
\label{i_rpn_top}
\vspace{-1.3em}
\end{figure}

The proposed IMF processor architecture shown in Fig. \ref{i_rpn_top} consists of an address event representation module (AER), a $128\times32$ bit asynchronous buffer, an IMF controller, and a $320\times 240$ SRAM macro.% ($W=320$, $H=240$ chosen to conform to QVGA resolution). 
The AER module follows an address event representation (AER) protocol~\cite{Boahen2004} to receive the event data asynchronously. It has a $10$-bit data (address and polarity of an event) line and two control lines- `Req' and `Ack' for synchronization and uses a $4$-phase handshake mechanism. %The AER module can be configured for a maximum of eight regions of exclusion (ROE)~\cite{Jyotibdha_EBBIOT} to ignore events that fall within those regions (e.g. to remove occluded regions from analysis).
The IMF has two clock domains: `sysClk' for filtering, and `aerClk' for processing raw events from an NVS. Since the asynchronous AER communication protocol takes multiple aerClk cycles to transfer an address of an event, the frequency of aerClk can be kept higher than the sysClk without overflowing the asynchronous buffer whose size can also be reduced as a result. The IMF controller performs three primary operations sequentially: clear memory, write memory, and filter image, as shown at the bottom of Fig. \ref{i_rpn_top}. The row and column signals from the IMF controller are $240$-b and $320$-b, respectively, following the dimensions of the macro except the bank select (BS). All signals are controlled either in parallel or independently to perform the three primary operations.

\begin{enumerate}
\item \textbf{Clear Memory}: Since the NVS only reports pixels with value $1$ which are generally sparse, the memory is fully cleared before writing $1$ at these few addresses. Bit-line (BL) and its complement (BLB) are driven to $0$ and $VDD$ respectively to clear the SRAM macro. Once a word line (WL) is made high, all SRAM cells in that particular row get cleared. Enabling a WL in every clock cycle stretches the execution time ($240$ clock cycles). In contrast, asserting all WLs high leads to a higher surge current at the bit-line pair (BL/BLB), which coupled with small size due to layout constraints may further trigger electromigration induced reliability issues. Furthermore, the strength and area of BL and BLB drivers need to be very high to clear all $240$ SRAM cells at once (worst case scenario). To address these concerns, we enable $16$ WLs in every clock cycle involving $240/16=15$ clock cycles to erase the memory.

\item \textbf{Write Memory}: Since the addresses of the events are non-contiguous by nature, unlike a conventional SRAM, either a single byte or word cannot be written. Consequently,  we implement a single bit writing circuitry enabling particular WL and bit-line pair (BL to $VDD$ and BLB to $0$ V) pointed by the yaddr and xaddr signals, respectively. We discuss the memory write and its associated circuitry in section ~\ref{nomf_hw_imp} in detail.
\item \textbf{Filter Image}: In general, background comprising ``0" valued pixels surrounds a noise pixel, and the noise should be removed by a filtering operation. To implement an $n\times n$ window of a filter, the IMF controller enables $n$ WLs and shorts $n$ BLs and BLBs separately using transmission gates ($n\in \{3,5\}$). Initially, both bit-lines (BL and BLB) get precharged to VDD. Once $n$ WLs are asserted high, one of the bit-lines discharges faster and causes bit flips. Next, we talk about the theory of the proposed filter and its circuit-level design.
\end{enumerate}

\begin{figure*}[t]
\centering
\includegraphics[scale=.63]{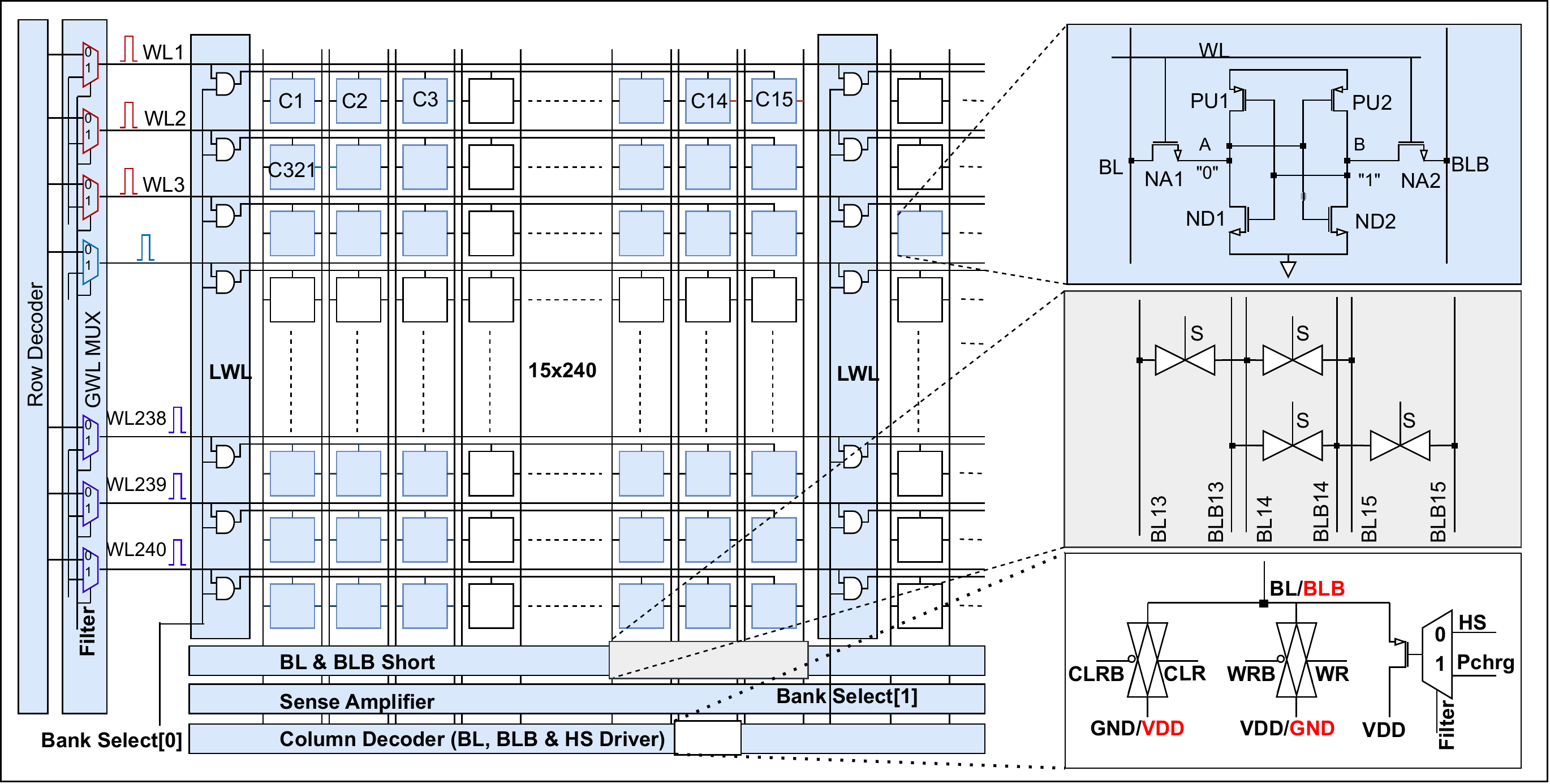}
\vspace{-0.5em}
\caption{Architecture of a $320\times240$ SRAM macro for image denoising. In filter mode, the IMF controller enables three consecutive word-lines (WL) to discharge bit-line (BL) and bit-line bar (BLB) simultaneously. BLs and BLBs of three successive columns are connected separately using transmission gates to implement a $3\times3$ patch/window, as shown in the inset (right-middle). This enables highly parallel noise filtering of $320\times3$ cells in two clock cycles.}% The circuitry at the right-bottom inset presents the BL/BLB driver for memory clear, write and filter operations.}
\label{filer_arch}
\vspace{-1.3em}
\end{figure*}

\subsection{NOMF: Hardware Implementation}
\label{nomf_hw_imp}
The proposed SRAM macro architecture shown in Fig. \ref{filer_arch} consists of $22$ separate $15\times240$ SRAM memory banks, local word line drivers (LWL), a global word line driver (GWL), a row and a column decoder. It supports QGVA as well as low-resolution images from an NVS such as~\cite{5648367},~\cite{Brandli:200837}. The  $320\times240$ SRAM macro is divided into $N_{bank}=22$ banks to minimize the dynamic bit-line power consumption~\cite{Chandrakasan:1995:LPD:560639} while writing an SRAM cell. Since the filter kernel sizes of $n=3$ and $5$ are supported, the number of columns per bank is kept equal to their least common multiple of $15$; this in turn leads to $N_{bank}=\ceil{320/15}=22$.  In SRAM write mode, global (GWL), and local word-line (LWL) blocks enable one of the word-lines (WL), and column decoder writes the data (``1") and its complement on the BL and BLB, respectively. The rest of the BLs and BLBs are precharged to VDD by the half select (HS) drivers to mitigate the read disturb issue of the half-selected cells in the selected bank, as shown in the bottom-right inset of Fig.~\ref{filer_arch}. %Fig. \ref{filer_arch} (bottom-right) illustrates the circuitry of BL/BLB driver for memory clear, write operations. The IMF controller erases all $320\times 240$ locations before writing a binary image frame. % The HS signal goes low for half-selected columns while writing the memory. %Similarly, while reading, RDB signal gets asserted to low for half a clock cycle to charge BL and BLB lines, and in the next half-cycle, P1 and buffer, U1 sense the BL or BLB value. 
%\begin{figure}[t]
%\centering
%\includegraphics[scale=0.6]{image/BL_DRIVE.eps}
%\caption{Peripheral circuitry for memory clear, write and read operations. The IMF controller erases all $320\time240$ locations before writing a binary image frame. HS signal goes low for half-selected columns while writing the memory. Similarly, while reading, the RDB signal gets asserted to low for half a clock cycle to charge BL and BLB lines, and P1 and buffer, U1 sense the BL or BLB value in the next half-cycle.}
%\label{bl_driver}
%\end{figure}
\subsection{Proposed Denoising Operation}
We leverage the read disturb phenomenon of a conventional $6$T-SRAM to implement the proposed NOMF. While writing an SRAM cell, the line driven to $0$ V, initiates the bit flip process in the SRAM cell. For instance, the SRAM cell shown in the top right of Fig. \ref{filer_arch} stores ``0". To write ``1'', BL and BLB are driven to VDD and $0$ V respectively. Once a WL line is made high, NA2 pulls down node B and triggers a bit flip procedure. While reading an SRAM cell, BL and BLB are charged to VDD in the first half of the read cycle, and in the next half-cycle, the bit lines are kept floating, and one of the WLs is asserted. As a consequence, one of BLs (BL or BLB) gets discharged due to the stored value in the SRAM. Likewise, having charged BL and BLB lines to VDD in a half cycle, if multiples WLs are enabled simultaneously in the next half keeping BL and BLB floating, one of the stored bits (``1'' or ``0") in the selected SRAM cells dominates and discharges one of the BL lines faster than other line similar to a read disturb situation. Consequently, the faster discharged line causes a bit flip to the SRAM cells. For instance, if the IMF controller enables three SRAM cells storing ``010", BL gets discharged faster than BLB. As a result, the SRAM cell storing ``1" gets flipped to ``0". At the end of this process, all the three SRAMs store ``0" indicating a majority operation.

We follow the above procedure of an SRAM read disturb to implement the NOMF for noise removal from an image since median filtering for a binary image is equivalent to the majority operation. The IMF controller enables $n$ consecutive word-lines (WL) to discharge BL and BLB simultaneously. BLs and BLBs of $n$ successive columns are connected separately using transmission gates to implement an $n\times n$ patch, as shown in Fig. \ref{filer_arch} for a $3\times3$ patch. The size of the sliding window can be configured as $n\times n$ shape where $n ~\in \{3,5\}$. However, with a larger value of $n=7$, the window size is comparable with the size of an object which results in higher approximation. The enable signals of the transmission gates for $3\times3$ filter are $110110110\cdots$ Similarly, for $5\times5$ implementation, the enable signals are $111101111011110\cdots$. These enable signals are active during the NOMF operation. Once NOMF is over, all these signals are made low.

%Besides, it demands larger SRAM cell area to meet the worst-case design criteria derived in Eq.~(\ref{mism}) later.  Therefore, we restrict values of $n$ to $3$ and $5$.   

\begin{figure}[t]
\centering
\includegraphics[scale=0.44]{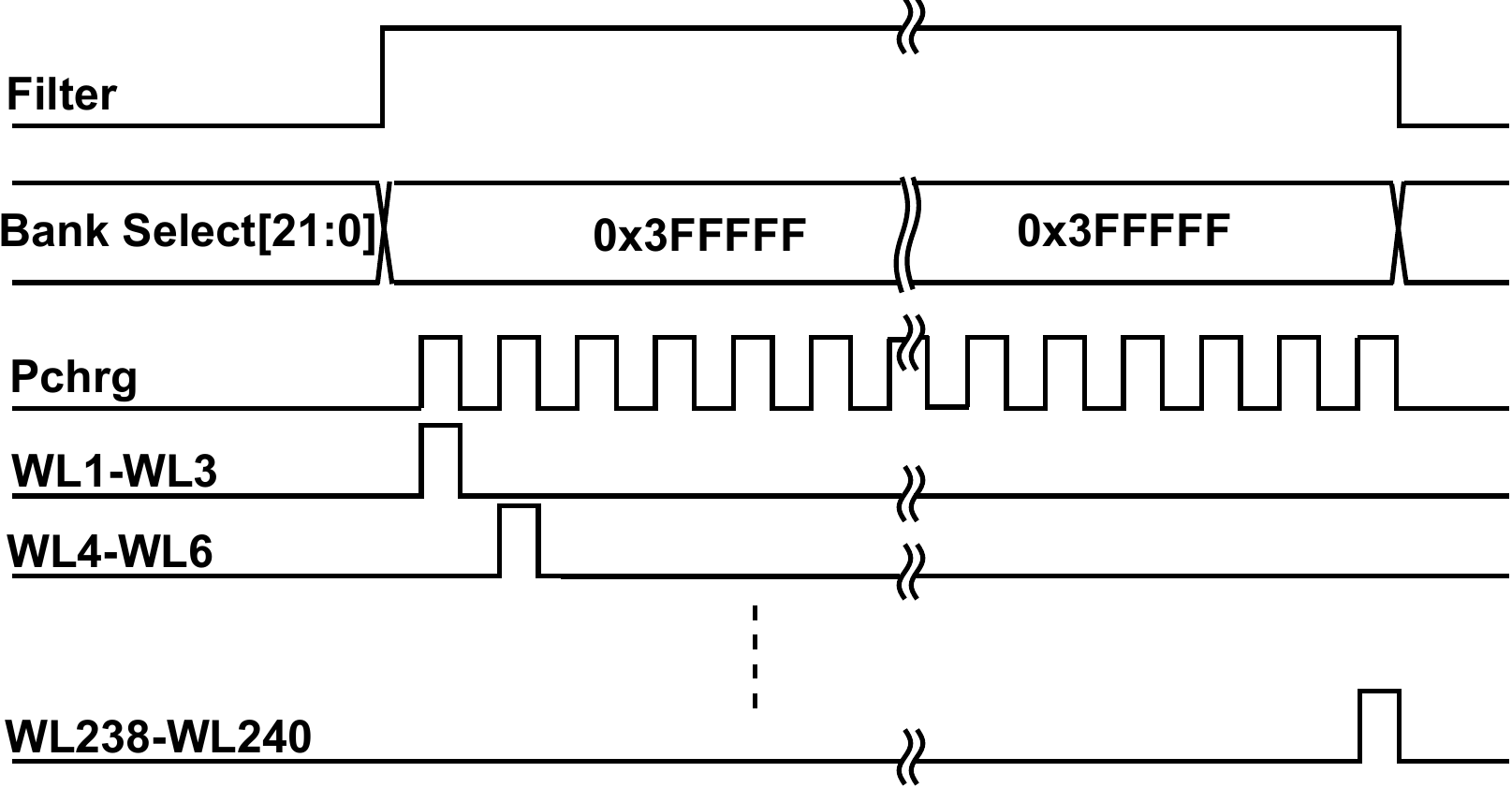}
\vspace{-0.5em}
\caption{Timing diagram of different signals in filter mode.}
\label{fig:timingdiagram}
\vspace{-1.2em}
\end{figure}

BLs and BLBs of the $n\times n$ cells are connected separately employing transmission gates, as shown in the right middle inset of Fig. \ref{filer_arch} for $n=3$. The signal $\text{S}$ is kept high throughout the filter operation. To avoid the lagging of discharge of BL and BLB lines, the resistance of the transmission gate, $R_{tg}$ is designed based on the following criterion:

\begin{equation}
R_{tg}C_{BL}<<C_{BL}\cdot\frac{VDD}{n\cdot i_s}
\label{eqn3}
\end{equation}
where $i_s$ denotes the discharging current of each SRAM cell, and $C_{BL}$ is a combination of the parasitic capacitor of metal routing of  BL or BLB and diffusion capacitor of access transistors ($\text{NA}_1$ or $\text{NA}_2$). Post-layout simulation after parasitic extraction provides $C_{BL}$ $\approx 140$fF. The IMF controller keeps all the bank select signals high, facilitating highly parallel processing in the memory. It processes $320\times n$ cells simultaneously in two clock cycles and repeats the procedure until all the rows are filtered requiring a total of $240/n$ repetitions of the above operation. At every alternate period, BLs and BLBs get precharged to VDD followed by the enabling of $n$ successive WLs in the next cycle (timing diagram is shown in Fig.~\ref{fig:timingdiagram}) causing ($n\times n$) SRAM cells to discharge BLs and BLBs simultaneously, as shown  in Fig.~\ref{bit_flip}(a). 
%Since there is a  difference in the number of ``0"s and ``1"s in a $n\times n$ kernel, one of the lines discharges and reaches $0$ V faster. This leads to bit flips of the minority pixels in the kernel (see Fig.~\ref{bit_flip}(a)). Notably, if the number of ``0"s is less than the number of ``1"s in a kernel, we refer ``0" as a minority pixel in that window and vice versa. 

Intuitively, the $n\times n$ patch can be thought of as a circuit where two latches of different strength and stored values are connected to BL and BLB.% Their strengths are determined by the number of ``0"s and ``1"s stored in the patch. 
 Whichever wins in discharging BL or BLB faster, imposes its stored value on the other. Next, we analyze the effect of non-idealities on this behavior.

\subsection{Effect of Statistical Variations}
\label{sec:variations}
In this section, we analyze the effect of statistical variations on the filtering operation. While discharging BL and BLB, a voltage difference between the lines builds up after WL is enabled and finally a bit is latched on the entire $n\times n$ patch when the trip point of the latch is reached. Denoting by $\Delta t$ the time difference between reaching of trip point on BL and BLB, we get:
\vspace{-0.2em}
\begin{align}
\vspace{-0.2em}
    \Delta t &= nC_{BL(B)}(\frac{V_{BL,trip}(k)}{I_{BL}(k)}-\frac{V_{BLB,trip}(k)}{I_{BLB}(k)})
    \vspace{-0.2em}
\end{align}
%nC_{BL}\frac{V_{BL,trip}}{I_{BL}}-nC_{BLB}\frac{V_{BLB,trip}}{I_{BLB}} \notag\\
 %   &\approx 
%\begin{equation}
%\Delta V=\frac{1}{n}\bigg(\frac{\Sigma i_0}{C_{BL}}-\frac{\Sigma i_1}{C_{BLB}}\Bigg)t
%\label{eqn4}
%\end{equation}
where $k$ denotes the number of non-zero (or ``1") pixels in the patch and we assume mismatch between capacitances on BL and BLB are negligible, trip points are denoted by $V_{BL(B),trip}(k)$ and discharge currents by $I_{BL(B)}(k)$ where the dependence on $k$ is made explicit. Each of these terms are also affected by statistical variations and this leads to a probability of error, $\epsilon$. To formalize this, we first note that ideally, we expect $\Delta t>0$ for $k\geqslant\ceil{\frac{n^2}{2}}$ and $\Delta t<0$ otherwise. Then, we can express the probability of error, $P(\epsilon)$ as follows:
\vspace{-0.2em}
\begin{align}
    P(\epsilon) &= P(\Delta t<0) \textit{ for } k\geqslant\ceil{\frac{n^2}{2}}\notag\\
    &+P(\Delta t>0) \textit{ for } k<\ceil{\frac{n^2}{2}}
\end{align}
 While we cannot obtain closed form solutions for these equations and need to resort to Monte Carlo simulations, we can gain some insight by analyzing extreme cases. For $k=0$ or $k=n^2$, one of $I_{BLB}$ or $I_{BL}$ equals $0$ and the mismatch does not affect the final decision resulting in $P(\epsilon)=0$. On the contrary, the NOMF takes the longest time to flip the minority pixels when the difference between the number of ``0"s and ``1" is one (i.e. $k=\floor{\frac{n^2}{2}}$ or $\ceil{\frac{n^2}{2}}$). This  deteriorates further due to the mismatch of the discharging current and capacitor and could force the majority pixels to flip wrongly. 

%where $\Sigma i_0$ and $\Sigma i_1$ represent the discharging current of BL and BLB due to the stored ``0"s and ``1"s in the $n\times n$ patch, respectively, and $t$ denotes time elapsed since comparison is enabled. In the best possible case, one of the currents is zero, and none of the bits flip.  This happens when all the bits in the patch are either ``0" or ``1". On the contrary, the NOMF takes the longest time to flip the minority pixels when the difference between the number of ``0"s and ``1" is one (worst case). This  deteriorates further due to the mismatch of the discharging current and capacitor and could force the majority pixels to flip wrongly. %A $1000$ points Monte-Carlo simulation in $65$ nm CMOS of BL and BLB discharging current of the above worst case scenario at VDD=$0.8$ V is shown in Fig. \ref{bit_flip}(b). The overlap region in the histogram is responsible for the unintended bit flips \textcolor{red}{and evaluates to a bit error ratio (BER) of 0.045}. 

\begin{figure}[t]
\centering
\includegraphics[scale=0.345]{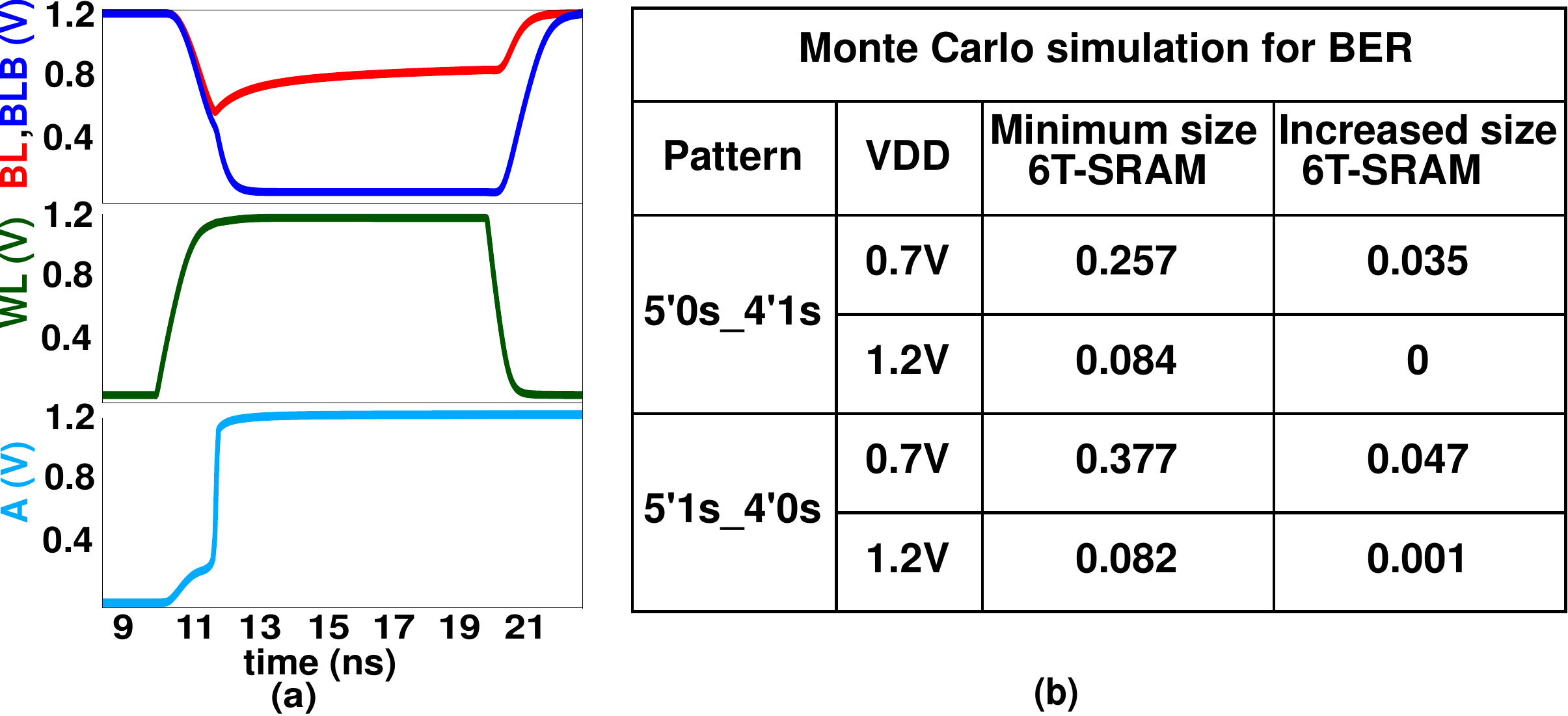}
\vspace{-0.5em}
\caption{(a) Bit flip of an SRAM memory cell in a $3\times3$ patch due to read-disturb. We initialize the $3\times3$ patch with $5``1"\text{s}\_4``0"\text{s}$ to simulate the NOMF in SPICE. BL and BLB are charged to VDD when the WL signal is made low. Initially,  node  A  stores ``0”, and  when WL goes high, BL and BLB start discharging. Since the number of ``1" is higher than that of ``0", BLB gets discharged faster than BL, and the stored value gets flipped at node A. (b) $1000$ points Monte-Carlo simulation at nominal temperature ($27^{\circ}$C) for BER.}%bit flip error ratio ($\mathrm{BER=\frac{\#~unwanted~bit~flips}{\#~MC~iterations}}$). at $5``1"\text{s}, 4``0"\text{s}$ and $4``1"\text{s}, 5``0"\text{s}$ scenarios (worst case).}% $2\times$ width and length of the access and pulldown NMOS of a $6$T SRAM resulted in an $\approx 8\times$ reduction of the number of unwanted bit flip compared to a minimum sized $6$T SRAM at $0.7$ V, and an error free operation at $1.2$ V.}
\label{bit_flip}
\vspace{-1.4em}
\end{figure}

However, the probability of  $\floor*{\frac{n^2}{2}}$ noise pixels appearing inside the faulty image patch is quite low. By analyzing the  dataset  described in section~\ref{imagede}, we observe that  the  probability  of  four  noisy  pixels occurring in  a  $3\times3$ image patch is only $\approx 0.001$ making the net bit flip error ratio (BER) quite small. The unintended bit flips reduce the object boundary when the majority pixel is ``1" and inserts a new object in the frame in the opposite scenario. Reduction of object boundary is not very critical since classification by CNN is not very sensitive to reduction of few boundary pixels; moreover, it can be corrected by a following region proposal block. The insertion of a new object in the frame is critical since it increases the processing in the subsequent stages but could potentially be corrected by a tracker operating on these regions across many frames\cite{singla2020ebbinnot}.

%Denoting the difference of discharge currents of BL and BLB by $\Delta I$ and bit-cell current as $I_b$, we can derive:
%\begin{align}
%\frac{\sigma}{\mu}(\Delta I) = %\frac{n}{|2k-n^2|}\frac{\sigma}{\mu}(I_b) \leq n\frac{\sigma}{\mu}(I_b)
%\label{mism}
%\end{align}
%where k represents the number of ``1'' in an $n\times n$ filter window, and the maximum value is achieved for $|2k-n^2|=1$. Hence, to reduce $\frac{\sigma}{\mu}(I_b)$, size of the bit-cell transistors and overdrive voltage $(V_{GS}-V_T)$ may be increased~\cite{Pelgrom}. In our case, width and length of NA1, NA2, ND1, and ND2 are increased by a factor of $2$ from its minimum value supported by the process, and low VT devices are used to maximize overdrive voltage. Hence, a trade-off exists between the choice of devices and leakage current.  
Nevertheless, the memory cell is designed to minimize bit errors due to variations. Intuitively, since BL and BLB are discharged by NMOS transistors with $I_b$ denoting unit cell discharge current, to reduce $\frac{\sigma}{\mu}(I_b)$, size of the bit-cell transistors and overdrive voltage $(V_{GS}-V_T)$ can be increased~\cite{Pelgrom}. In our case, width and length of NA1, NA2, ND1, and ND2 are increased by a factor of $2$ from its minimum value supported by the process (increased W=$240$nm, L=$120$nm), and low VT devices are used to maximize overdrive voltage. Fig. \ref{bit_flip}(b) tabulates the results of Monte Carlo simulations at VDD = $0.7$ V and $1.2$ V, respectively, for a minimum sized SRAM cell and the redesigned one, showing almost error free operation at $1.2$ V and $\approx 8$X reduction in BER ($\mathrm{BER=\frac{\#~unwanted~bit~flips}{\#~MC~iterations}}$) at $0.7$ V. %Note that this sizing results in a slightly lower trip point but this is not corrected by increasing PMOS size to minimize extra area overhead.

\begin{figure}[t]
\centering
\includegraphics[scale=0.45]{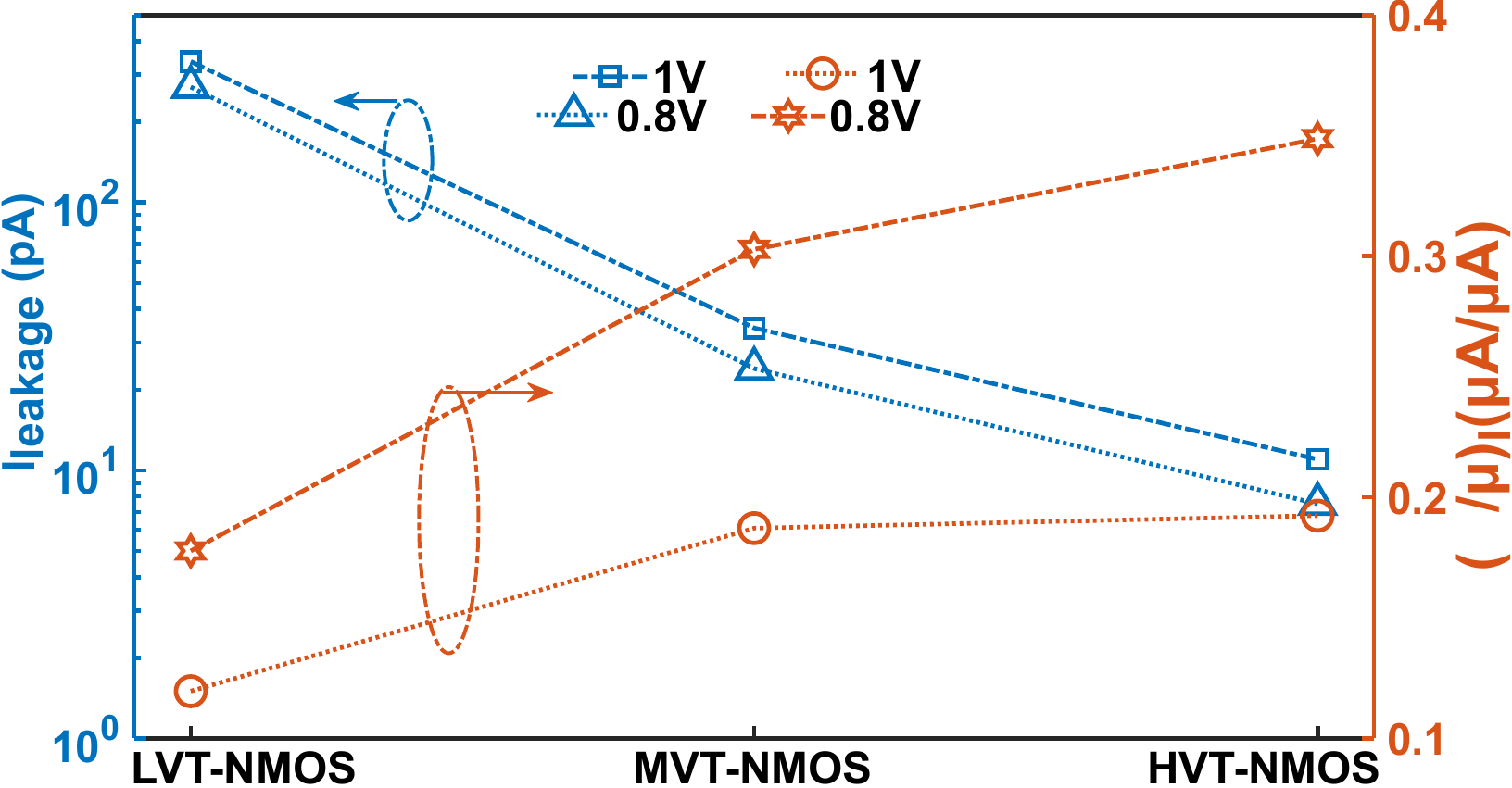}
\vspace{-0.5em}
\caption{Leakage current and normalized standard deviation of BL/BLB discharge current of an SRAM cell for low-VT (LVT), medium-VT (MVT), and high-VT (HVT) NMOS. The discharge current variation is significant at lower supply voltage and higher VT devices due to lower overdrive.}
\label{tradeoff}
\vspace{-1.0em}
\end{figure}

Since we are using the low-VT (LVT) devices, a trade-off exists between the choice of devices and leakage current. To explore this, we have simulated the leakage current, and normalized standard deviation of BL/BLB discharge current ($(\frac{\sigma}{\mu})_I$) of an SRAM cell for low-VT (LVT), medium-VT (MVT), and high-VT (HVT) NMOS at $0.8V$ and $1.0V$ as shown in Fig. \ref{tradeoff}. We can see that LVT NMOS provides better tolerance towards mismatch at a fixed supply voltage due to the higher overdrive at the cost of higher leakage current. However, we can shut down the memory once processing ends to minimize the extra power dissipated due to leakage currents. It is worth mentioning that the increased-size $6$T-SRAM has $45\%$ higher area and $1.57\times$ higher parasitic capacitance on BL and BLB than the minimum size $6$T-SRAM, leading to a higher dynamic power during operation. However, the increased-size $6$T-SRAM enables more voltage scaling at the same error rate. Therefore, it can be calculated that the increased-size $6$T-SRAM consumes $\approx3X$ lower power than the  minimum size $6$T-SRAM at iso-accuracy (BER) conditions.

%The final coefficient of variation ($\frac{\sigma}{\mu}$) of the BL/BLB discharge current is $\approx 0.12$ at $1V$, as shown in Fig. \ref{tradeoff}. 

\begin{figure}[t]
\centering
\includegraphics[scale=0.55]{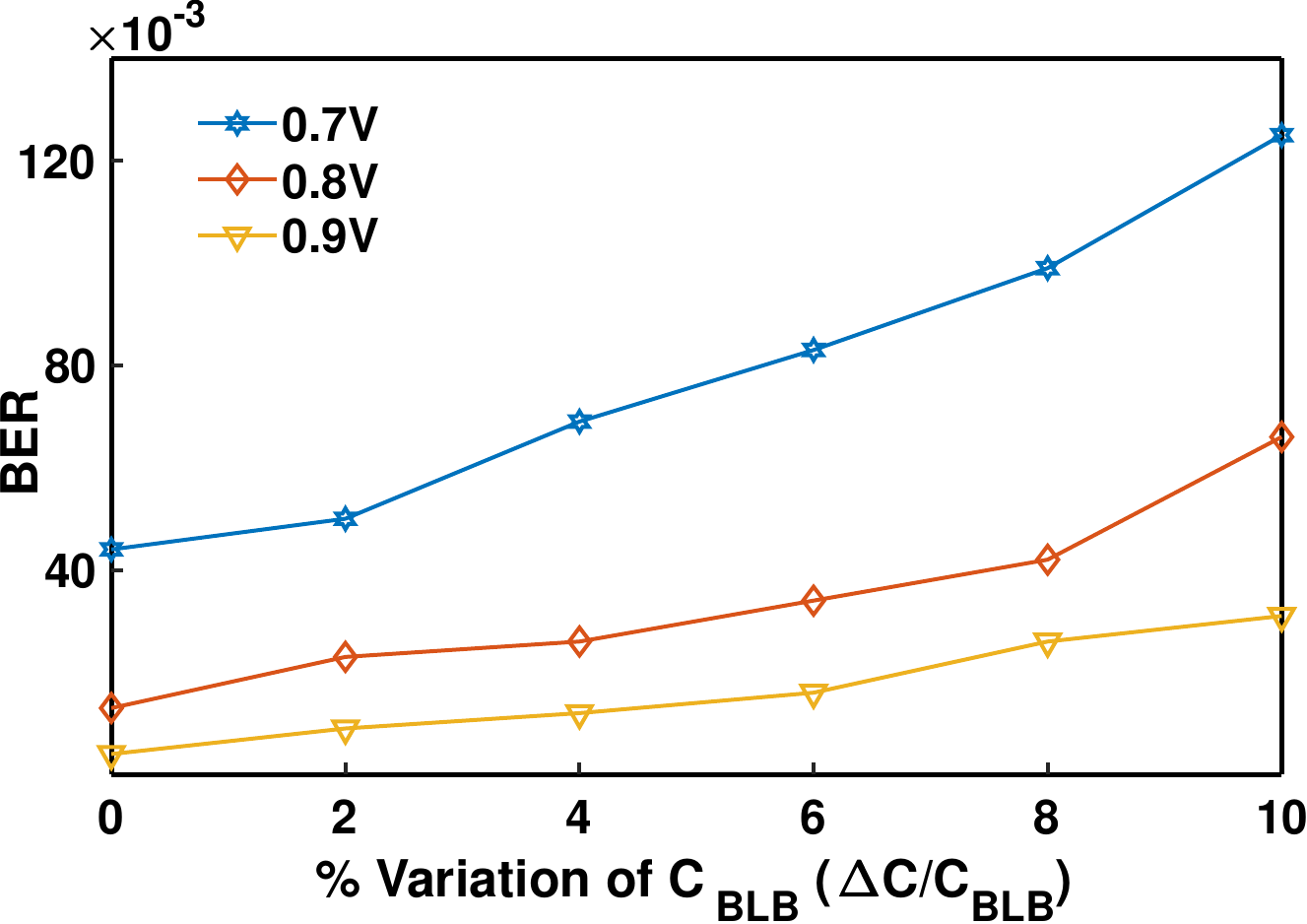}
\vspace{-0.5em}
\caption{Effect of unbalanced parasitic capacitance on BER, which decreases at higher supply voltages.}
\label{capvariation}
\vspace{-1.0em}
\end{figure}

In a post-layout simulation, we have observed that the parasitic capacitance on the BL and BLB are almost the same ($139.6$fF and $140.4$fF). However, due to inaccuracies of fabrication steps, there can be a difference in parasitic capacitance ($\Delta C$) between BL and BLB. To see the effect of unbalanced capacitance, we have simulated $1000$ points Monte Carlo simulation at the worst-case pattern ($5$``1”s\_$4$``0”s) and several supply voltages for different BLB capacitance values. The BER is plotted in Fig.~\ref{capvariation}, where the x-axis represents the percentage variation of BLB capacitance. The variation of the parasitic capacitance ($\Delta C$) in opposite the direction reduces the BER for $5$``1”s\_$4$``0”s pattern. Similar analysis can be performed for $4$``1”s\_$5$``0”s pattern for different BL capacitance values. The capacitance mismatch can be partially corrected by connecting small capacitors ($~\sim5-10$fF) on both BL and BLB lines and enabling one of them if required.

\vspace{-1.5em}
\subsection{Effect of Temperature variations}
\label{sec:temperature}
\begin{figure}[b]
\centering
\includegraphics[scale=0.401]{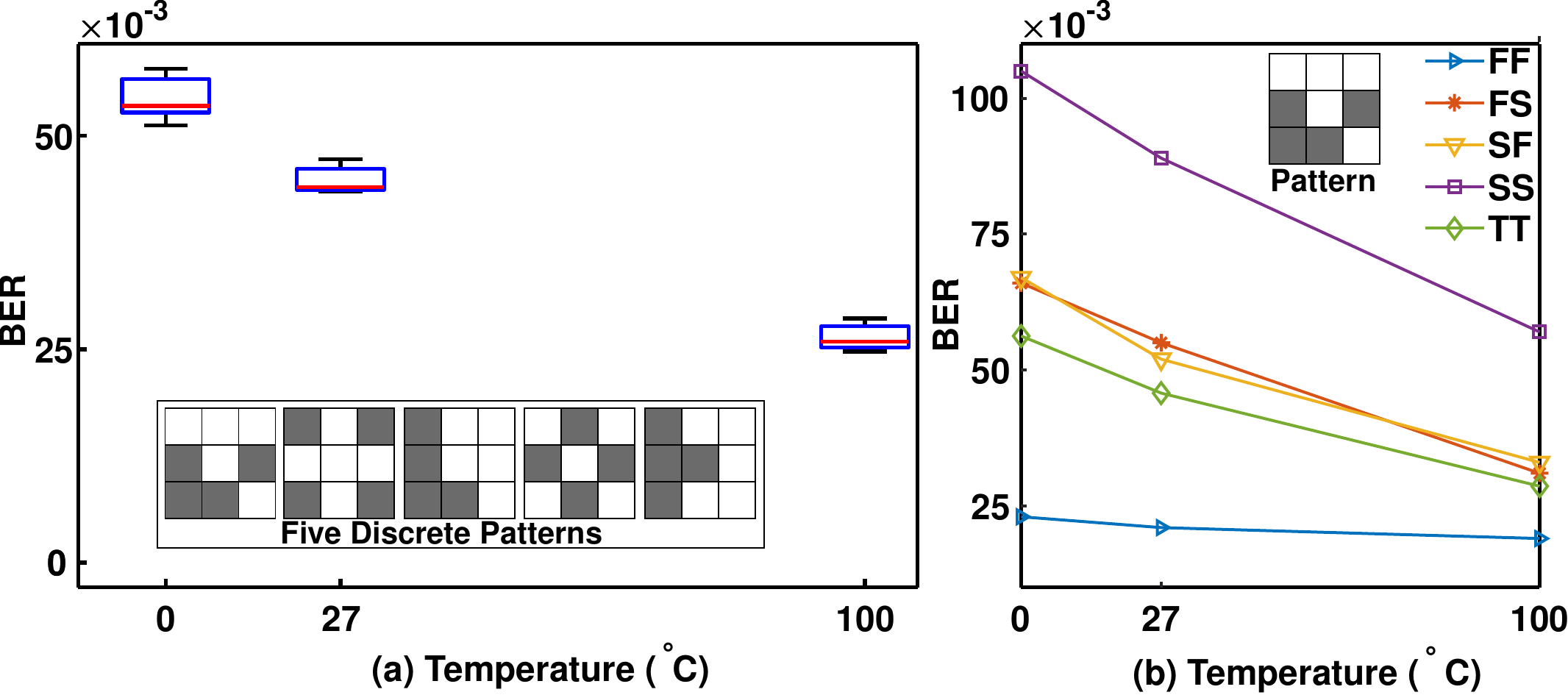}
\vspace{-0.5em}
\caption{(a) Monte Carlo simulation of a $3\times3$ patch initialized at five random discrete $5$``1”s\_$4$``0”s patterns across temperature at $0.7V$ with $8000$ points for each simulation. The white and dark colors in the patterns (at the inset) represent ``1”s and ``0”s stored in a $3\times3$ SRAM window, respectively. Bit flip error ratio ($\mathrm{BER=\frac{\#~unwanted~bit~flips}{\#~MC~iterations}}$) reduces at higher temperature. (b) $1000$ points Monte Carlo simulation of a $3\times3$ patch across different corners at $0.7V$. BER increases at SS corner due to the lower overdrive of the transistors.}
\label{bittemp}
\vspace{-1.2em}
\end{figure}

To evaluate the robustness of the proposed NOMF across temperature, we have performed $8000$ point Monte Carlo (MC) simulations of a $3\times 3$ image patch initialized at five random discrete $5$``1”s\_$4$``0”s patterns (inset of Fig.~\ref{bittemp}(a)) chosen randomly out of $9 \choose 4$ possible patterns. Fig. \ref{bittemp}(a) presents the BER ($\mathrm{BER=\frac{\#~unwanted~bit~flips}{\#~MC~iterations}}$) of an SRAM with $3\times 3$ filter window at different temperatures and VDD $=0.7V$. The Fig. \ref{bittemp}(a) illustrates that with an increase in temperature, the number of unwanted bit flips reduces. This is reasonable since at a higher temperature, the threshold voltage of an NMOS reduces, which in turn lowers the normalized  standard  deviation  of  BL/BLB discharge current ($\frac{\sigma}{\mu}(I_b)$). The increased error at low-temperature can be nullified by increasing the supply voltage/WL drive linearly with lower temperatures~\cite{volttemp}. Fig. \ref{bittemp}(b) presents the $1000$ points Monte Carlo simulation across different corners. As expected, the BER increases at SS corner due to the lower overdrive of the transistors. 
%Note that the simulated number of unintentional bit flip is higher than that of the measured values. 
\vspace{-0.5em}
\begin{table}[t]
\centering
\caption{Comparison of different filters for an image of size, $\text{M}$=$\text{W}\times\text{H}$}
\vspace{-0.5em}
 \begin{threeparttable}
%\resizebox{\columnwidth}{!}{
    \begin{tabular} {|M{1.57cm}|M{0.60cm}|M{0.94cm}|M{0.94cm}|M{1.2cm}|M{0.65cm}|}\hline
          & Input &\# memory read &\# memory write &\# operations &\# SRAM Cells \\ \hline
        NN-filt &Events &$\beta_t\gamma n^2\text{M}\tnote{}$ &$\beta_t\gamma \text{M}$  &$\gamma n^2\text{M}$ &$\beta_t \text{M}$\\ \hline
        Median Filter &EBBI	 &$n^2\text{M}$  &$\text{M}$  &$n^2\text{M}$ &$2\text{M}$\\ \hline
        NOMF &EBBI&$\text{M}$  &$\text{M}$ &$\text{M}$ &$\text{M}$\\ \hline
        NOMF+IMC &EBBI &$\text{M}/n$  &$\alpha\text{M}$ &$0$ &$\text{M}$\\ \hline
    \end{tabular}
% }
\label{tab2}
\begin{tablenotes}
   \item[] $\beta_t=16$, $\gamma\approx12.7\%$, $\text{M}=\text{HW}$, $\alpha\approx1.5\%$, $n\in \{3,5\}$.  
  \end{tablenotes}
  \end{threeparttable}
\end{table}

\subsection{Performance}
\label{performance}
The advantages of the proposed in-memory computing-based NOMF are four-fold. Firstly, it reduces the dynamic power consumption while reading SRAM cells for filtering by a factor of $n$. The proposed approach charges $n$ BL and BLB lines once to filter out $n^2$ pixels, whereas the traditional method requires to charge $n^2$ BL and BLB lines to read the same number of pixels. Secondly, it eliminates the usage of a sense amplifier to detect a voltage difference between BL and BLB lines. The $n\times n$ SRAM cells act as a sense amplifier and decides on the basis of the majority bit. Thirdly, the proposed approach does not consume any BL and BLB dynamic power during writing the filtered value to the $n\times n$ SRAM cells since the discharges of BL and BLB are related to the read operation. Lastly, it consumes minimal energy to flip the minority pixels (only noise and boundary pixels of an object).

A comparison of the proposed in-memory computing-based NOMF with other event and frame based denoising techniques are shown in Table \ref{tab2} for processing a $W\times H$ image. The event-based nearest neighbour filter (NN-filt)~\cite{Gonzalez:2006:DIP:1076432} stores the timestamp of an incoming event using $\beta_t$ ($\beta_t$=16) bit per timestamp~\cite{Jyotibdha_EBBIOT}. Further, it marks the event as valid if the difference of timestamps in an $n\times n$ spatial neighbourhood is less than a specified threshold. The parameter, $\gamma$ represents the average number of events during the frame duration, which can be estimated as the average object size (${\leq}0.7\%$ (bike)-$12\%$ (truck) --given in Table~\ref{tabdata2}) times the average firing rate ($\approx2$) of a single pixel. As discussed, IMC reduces the number of memory read while filtering images by a factor of $n$. Parameter $\alpha$ represents the fraction of the pixels that need to be flipped for the filter implementation (only noise and boundary pixels of the objects). We estimate ${<}\alpha{>}\approx 0.015$ by analyzing $300k$ image frames from a DAVIS camera recording traffic scenes\cite{singla2020ebbinnot} resulting in $<135\times$ memory writes for our proposed system compared to NN-filt.

Compared to digital implementations of both median filter and NOMF, the proposed IMC based NOMF has at least $n$ times less memory reads and $\frac{1}{\alpha}$ times less memory writes. Furthermore, no operations by an external processor are needed in the proposed NOMF since all operations are done within the memory.
\section{Results}
\label{hwresults}
\begin{figure}[t]
\centering
\includegraphics[scale=0.46]{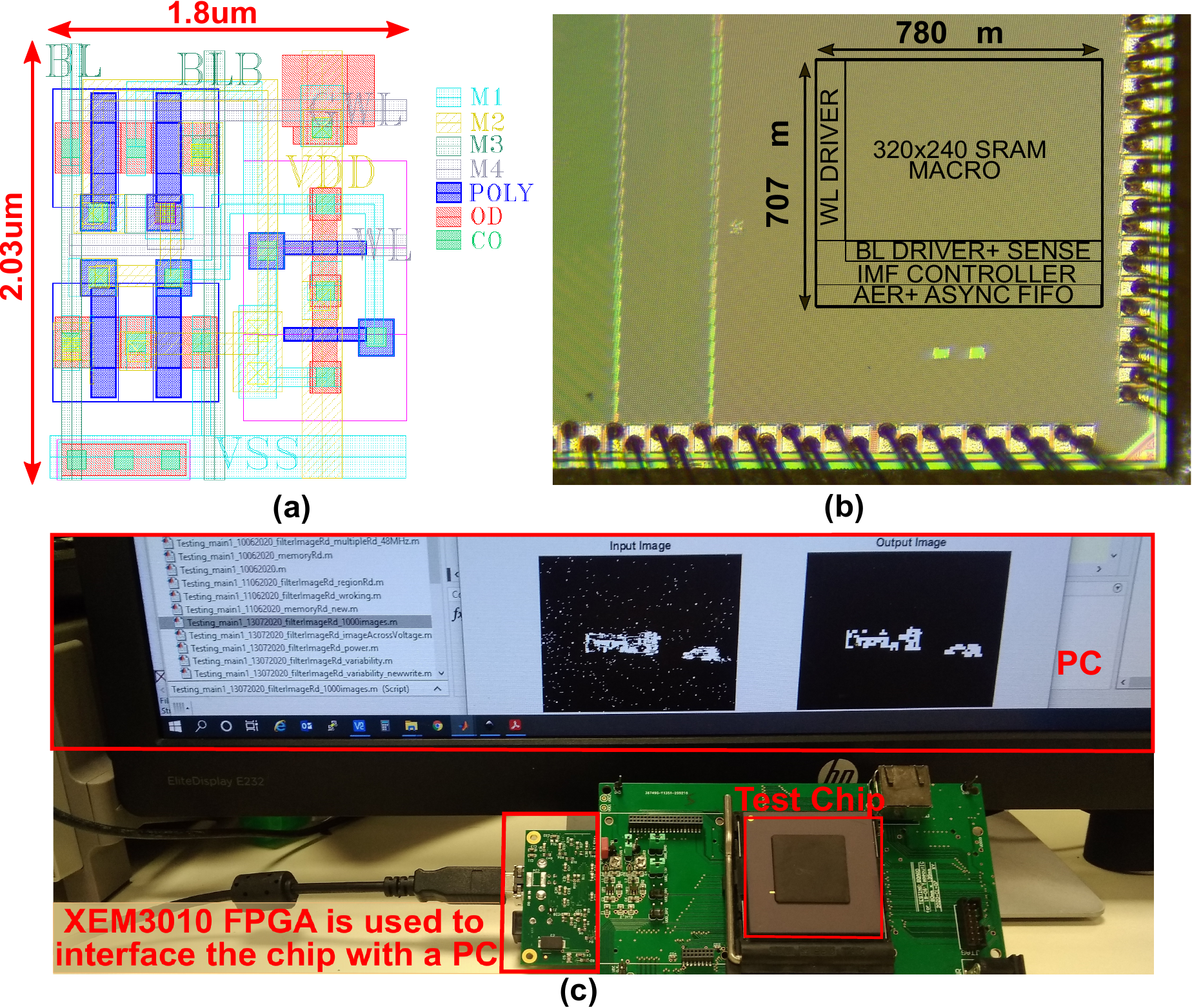}
\vspace{-0.8em}
\caption{(a) SRAM cell layout (b) chip micrograph with $320\times 240$ memory macro and on-chip modules (AER, FIFO, IMF controller, memory peripheral) for memory write and image denoising. (c) Test setup: external XEM3010 FPGA board from Opal Kelly is used to interface the chip with a PC. The PC sends the noisy image to the chip and receives the filtered image through the XEM3010 FPGA board.}
\label{layout}
\vspace{-1.2em}
\end{figure}
We implement the proposed NOMF filter in a $65$nm LP CMOS process. The unit SRAM cell layout and micrograph image of the prototype chip are shown in Fig. \ref{layout}(a) and (b) respectively. The prototype occupies $0.55mm^2$ and has utilization of $78.6\%$ and a capacity of $75$kb. We do not include the AER and asynchronous FIFO areas in the calculation of area utilization since they are outside the macro. An Opal Kelly FPGA is used to interface the PC with the test chip as shown in Fig. \ref{layout}(c) and is responsible for sending noisy images to the chip and relaying filtered images back to the PC for further analysis. In the following sections, we present results of characterization of the designed IC in detail and compare the performance of the proposed NOMF with the conventional median filter in the applications of object tracking and detection.

\subsection{Mismatch Characterization}
In this section, we measure the effect of unit cell current and latch trip point variation on the filtered image. Since the memory has a capacity of $75$ kb, the whole memory can be regarded as $8480$ $3\times 3$ image patches. The $n\times n$ patches in the memory are initialized with particular patterns of $k$ ``1"s and $(n^2-k)$ ``0"s where $k \in \{3,4,5,6\}$ and $n=3$. The IMF controller enables the denoising operation once the initialization is over. Eventually, the external FPGA board (shown in Fig.~\ref{layout}(c)) is used to read out the filtered image and detect the unintentional bit flip due to mismatches of the $n^2$ SRAM cells. We carry out the same experiment for ${n^2\choose k} $ discrete patterns of $k$ ``1"s  and $(n^2-k)$ ``0"s initialized in all patches. 

\begin{figure}[t]
\centering
\includegraphics[scale=0.415]{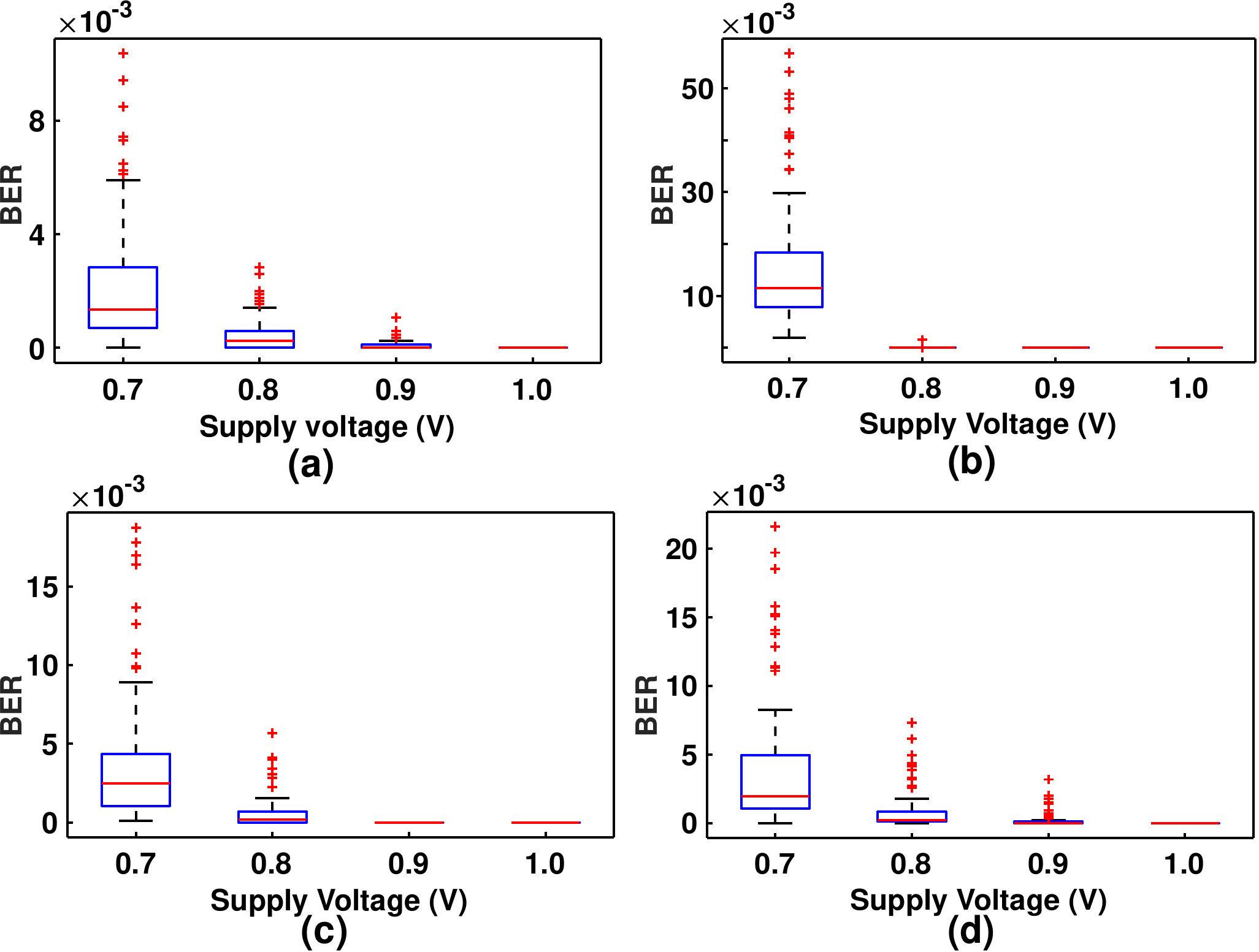}
\vspace{-0.5em}
\caption{Bit flip error ratio ($\mathrm{BER=\frac{\#~unwanted~bit~flips}{\#~patches}}$) of four devices due to the discharge current mismatch of nine SRAM cells of a $3\times3$ filter patch where $\mathrm{\#~patches=8480}$. The measurement is carried out for 126 $5$``1”s\_$4$``0”s discrete patterns across supply voltages at nominal temperature.}
\label{boxplot}
\vspace{-1.3em}
\end{figure}

Fig. \ref{boxplot} shows the BER ($\mathrm{BER=\frac{\#~unwanted~bit~flips}{\#~patches}}$) of four chips across supply voltages at $k=5$ ($126$ discrete patterns of $5$``1”s\_$4$``0”s), and nominal temperature where $\mathrm{\#~patches=8480}$. Note that we also observe the unintentional bit flip in two devices for $k=4$ at VDD=$0.7V$ and $0.8V$. However, the number of bit flips at $k=4$ is lower than that at $k=5$. It is also worth mentioning that there is no unintentional bit flip at $1V$, as shown in Fig. \ref{boxplot}. %It also validates the design criteria, as mentioned in Eq. (\ref{mism}).
Note that the unintentional bit flip at $k=5$ reduces the object boundary.

\begin{table}[h]
\centering
\caption{Mean object sizes (height$\times$width)~\cite{singla2020ebbinnot}}
\vspace{-0.5em}
%\resizebox{\columnwidth}{!}{
    \begin{tabular} {|M{1.3cm}|M{1.3cm}|M{1.3cm}|M{1.3cm}|M{1.3cm}|}\hline
          
Location & Car/Van &Bus &Bike &Truck \\ \hline
       Location 1 &$16\times42$ &$31\times94$ &$15\times21$ &$22\times50$\\ \hline
        Location 2 &$25\times47$ &$52\times107$ &$17\times22$ &$35\times61$\\ \hline
        Location 3 &$34\times82$ &$64\times180$ &$26\times44$ &$50\times104$ \\
        \hline
    \end{tabular}
\label{tabdata2}
\vspace{-1.2em}
\end{table}

\begin{figure}[t]
\centering
\includegraphics[scale=0.37]{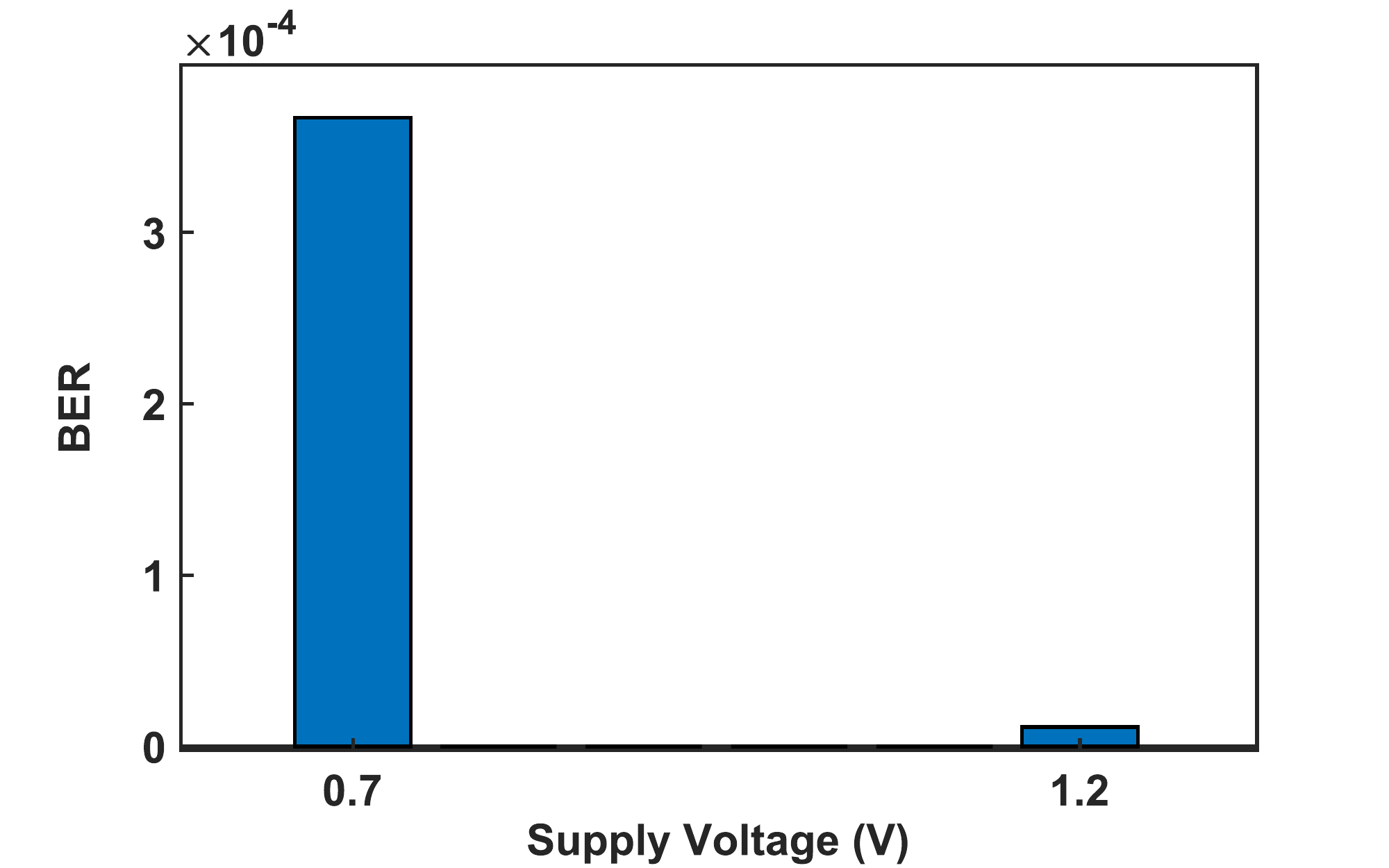}
\vspace{-0.5em}
\caption{Mean BER of the IMC based NOMF measured for $1000$ noisy images at VDD = $0.7V$ and $1.2V$.}
\label{histo}
\vspace{-1.0em}
\end{figure}
\begin{figure}[t]
\centering
\includegraphics[scale=0.5]{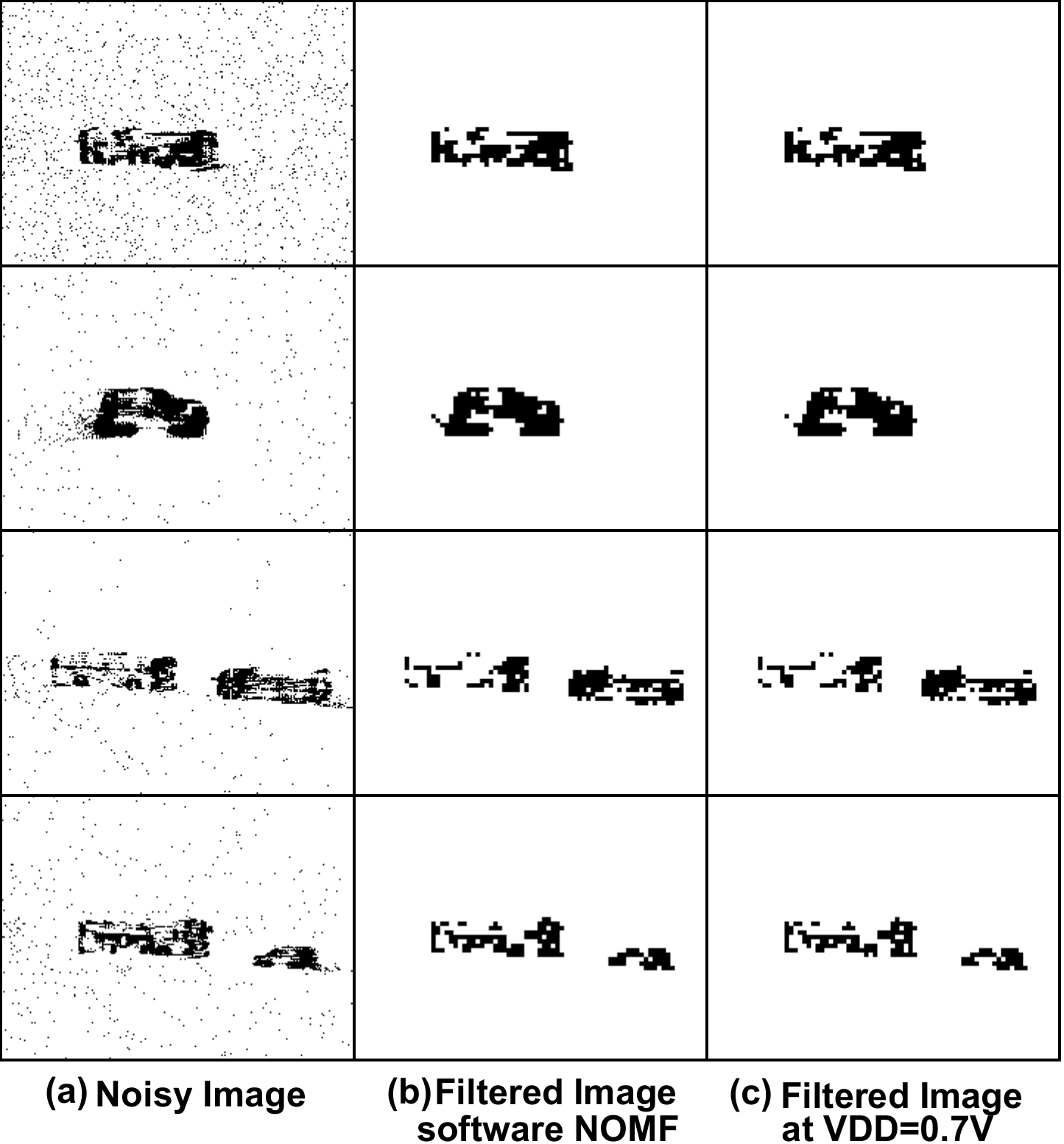}
\vspace{-0.5em}
\caption{(a) Noisy image (b) Filtered image: software NOMF operation (c) Filtered image at $VDD=0.7$V. There are $3,7,9$, and $6$ errors in the filter images from right top to bottom at $VDD=0.7V$, respectively. However, the impact of the error on the objects' shape and the boundary is insignificant.}
\label{filterimages}
\vspace{-1.3em}
\end{figure}

\subsection{Image Denoising}
\label{imagede}
To show the effect of the unintentional bit flip in real video recordings, we fed $1000$ event-based binary images (EBBI) from the traffic dataset\cite{singla2020ebbinnot}\footnote{Dataset: https://zenodo.org/record/3839231} captured using a DAVIS camera~\cite{Brandli:200837}. We describe the dataset in brief for completeness. A DAVIS setup was deployed at three places to collect three recordings of varying duration at different lens settings. Subsequently, events are aggregated at a frame rate of $15$Hz ($t_f = 66 $ms) to create event-based binary image (EBBI) frames. Table \ref{tabdata2} summarises  the sizes of different objects at three different locations. We randomly select $1000$ frames containing objects from the three recordings to validate the performance of NOMF. Note that each image dimension is $240\times180$.

The external FPGA was used to feed in  the images and read out the filtered image ($I_H$). Subsequently, we subtract the filtered images from the images ($I_S$) filtered following the software NOMF operation to calculate the bit error ratio (BER) as shown in Eq. (\ref{error}): 
\vspace{-0.2em}
\begin{align}
BER = \frac{1}{whN}\sum\limits_{k=1}^N\sum\limits_{i=1}^w\sum\limits_{j=1}^h |I_H^k(i,j)-I_S^k(i,j)|
\label{error}
\end{align}
where $w$ and $h$ represent the width and height of the image respectively and $N$ denotes the number of images considered. Fig. \ref{histo} presents the BER for $1000$ images at $VDD=1.2$V and $VDD=0.7$V. As expected, the BER at $VDD=1.2$V is $\approx 10^{-5}$, much lower than that at $VDD=0.7$V ($\approx 3.6\times 10^{-4}$). Furthermore, since the probability of occurrence of critical patterns ($5$``1”s\_$4$``0”s and $4$``1”s\_$5$``0”s patterns) in a real image frame is low, the BER in this case is approximately one order of magnitude lower than the BER at $0.7$V in Fig. \ref{boxplot} where each image patch was initialized to a critical pattern. 
%Also, the BER for real images is a bit higher than the BER obtained in individual cell characterization in Fig. \ref{boxplot}.  A potential reason is that the  access transistors, NA1, and NA2 of Fig. \ref{filer_arch} of an unselected SRAM, contributes different capacitances  ($\Delta C\approx0.017$fF) to the BL and BLB due to the different voltages across them. Since most of the SRAM cells along a column for real images contains ``0", a SPICE simulation shows that BLB experiences $\approx4.08$fF higher capacitance than the BL line. This systematic offset ($\approx 3\%$) is potentially responsible for the error at higher voltages. The contribution of the leakage current of the access transistors of an unselected SRAM cell is negligible, and hence, the systematic offset due to different capacitances can be corrected by adding offset deliberately in each column. It is worth mentioning that, since we initialize the $\approx 50\%$ of the SRAM cells in a column with ``1" while testing a single $3\times3$ filter window, we do not encounter any unintentional bit flip above $VDD=1V$ in Fig. \ref{boxplot}.

Fig. \ref{filterimages} shows samples of the (a) noisy images, (b) filtered images following the software NOMF operations, and (c) filtered images deploying the designed chip at $VDD=0.7V$. The filtered images from right top to bottom at $VDD=0.7V$ experience $3,7,9$, and $6$ errors, respectively. However, it can be seen from Fig. \ref{filterimages} that the impact of the error on the objects' shape and the boundary seems insignificant. In the next section, we will evaluate the effect of these errors on region proposal and tracking which are the following blocks in the signal processing chain.

\begin{figure}[t]
\centering
\includegraphics[scale=0.38]{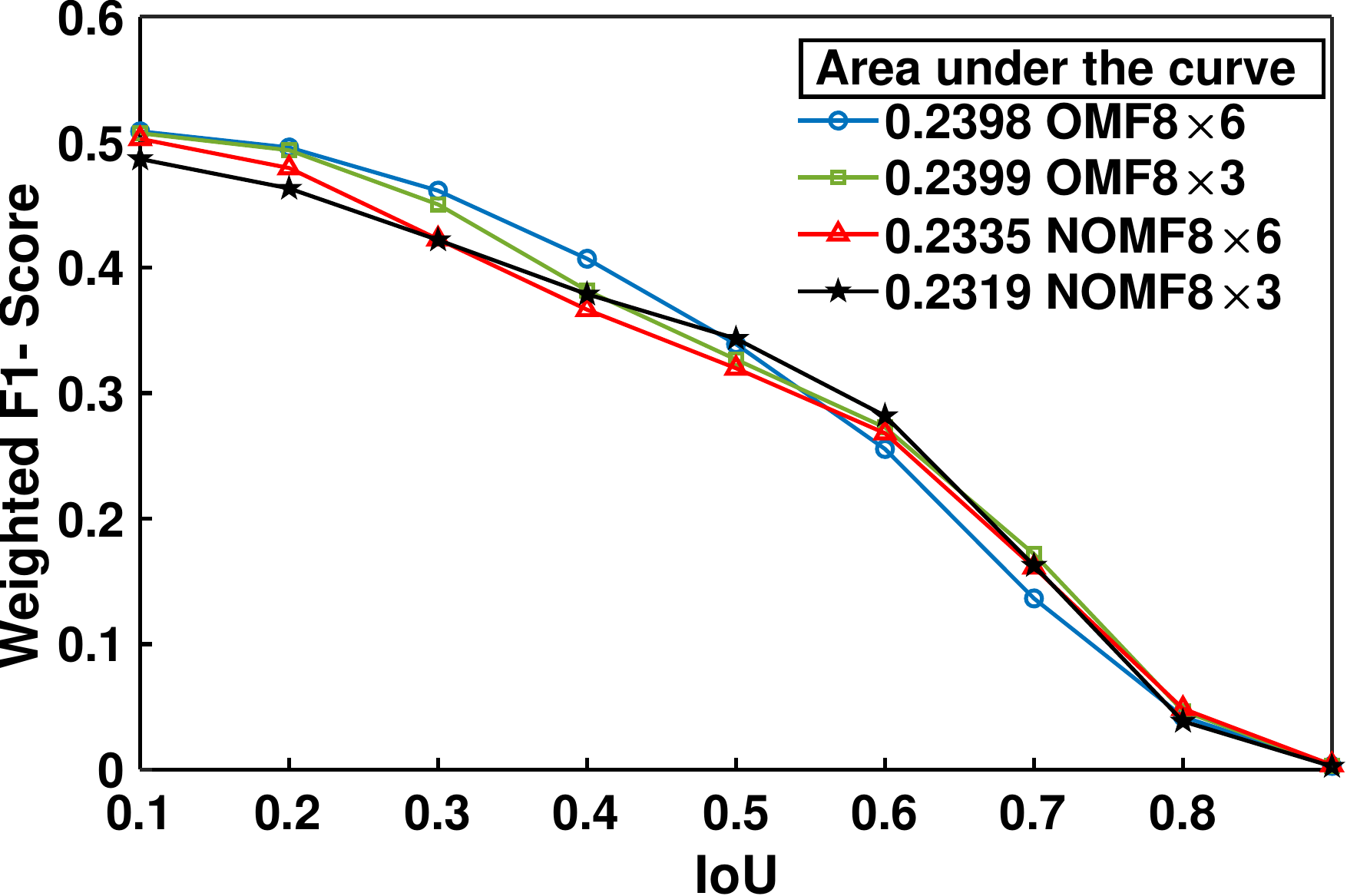}
\vspace{-0.5em}
\caption{Weighted F1-Score of an overlap based tracker (OT) using  filtered images following conventional median filter (OMF- overlap median filter) and NOMF for different IoU values.}
\label{f1score}
\vspace{-1.2em}
\end{figure}

\subsection{System level evaluation}
\subsubsection{Object Tracking}
To evaluate the effect of NOMF filtered images at the system level and compare with the conventional median filter, we feed both of them $\approx 70k$ raw noisy images from the earlier dataset and the filtered images are sent as input to the object tracking pipeline described in~\cite{singla2020ebbinnot}. Since data transfer between the PC and the IC is a bottleneck in terms of speed and there is little differences between the chip output and software NOMF operation (average $1.7$ and $0.048$ errors per image at $0.7$V and $1.2V$ respectively as shown in Fig.~\ref{histo}), we perform this system evaluation entirely in software. Moreover, the error occurs at the boundary of the object, as shown in Fig.~\ref{filterimages}--this gets mitigated by scaling down the image as part of the region proposal\cite{singla2020ebbinnot}. 

Note that the tracking involves region proposal network, and we follow connected component labeling (CCL)~\cite{HE201725} algorithm to estimate the region of interests (ROIs) followed by an overlap tracker (OT)\cite{singla2020ebbinnot} which is a simplified form of Kalman filter. In order to evaluate the system performance, we calculated the weighted $F1-score$, $F1^{wgt}_{thr}$ using all tracks predicted for all test recordings as follows: 
\begin{align}
&IoU =\frac{A_{GT} \cap A_P}{A_{GT}\cup A_P} \label{eqnfa}\\
&Precision, P^i_{thr}=\frac{true ~positive~ regions}{total~ proposed ~regions} \label{eqnfb}\\
&Recall, R^i_{thr}=\frac{true~ positive~ regions}{total ~ground ~truth ~regions} \label{eqnfc}\\
&F1-score, F1^i_{thr}=\frac{2\times P^i_{thr}\times R^i_{thr}}{P^i_{thr}+R^i_{thr}} \label{eqnfd}\\
&F1^{wgt}_{thr}=\frac{\sum\limits_{i=1}^rN^i_{tracks}\times F1^i_{thr}}{\sum\limits_{i=1}^rN^i_{tracks}}\label{eqnfe}
\end{align}
where $A_{GT}$ and $A_P$ denote the area of manually annotated ground truth and region proposed by the OT encapsulating an object, respectively. If the IoU of a proposed region is greater than a threshold, $thr$ the region is assumed to be a true positive region. $N^i_{tracks}$ denotes the number of tracks in recording $i$ and $1\leq i\leq r$, where $r$ represents the total number of recordings. We use the $F1-score$ as a metric that combines both precision and recall to compare different filtering methods.

Fig. ~\ref{f1score} presents a MATLAB simulation of the weighted $F1-score$  of an overlap based tracker (OT) using  filtered images following conventional median filter and NOMF for different IoU thresholds ranging from $0.1$ to $0.9$. This allows us to evaluate the performance of the system over a range of thresholds and the area under this curve (AUC) is a good metric to compare robustness and performance of all methods.
Following the method in \cite{singla2020ebbinnot}, we scale down the images for the CCL algorithm used in the analysis as follows
\vspace{-0.2em}
\begin{align}
\vspace{-0.2em}
I^{a,b}(i,j)&=\lor_{k=0,l=0}^{k=a-1,l=b-1}I(ia+k,jb+l)\label{eqndsa}
\vspace{-0.2em}
\end{align}
% &i<\floor*{w/a},j<\floor*{h/b}
where $I(i,j)\in\{0,1\}$ and $a,b$ are rescaling factors along the horizontal and vertical axis, respectively, and $\lor$ represents the logical-OR operation on $a\times b$ window. $w$ and $h$ denote the width and height of an image frame, respectively. We noticed that the rescaling of images in CCL majorly helps in handling the fragmentation observed in the recordings at the original resolution. This fragmentation occurs due to the invariance in illumination on the glass windows of the objects. Hence, an appropriately chosen downscaling may be able to merge the fragmented objects such as buses, trucks, etc. 

It is essential for the region proposal block to correctly identify the regions for the tracker to run properly and to improve the F1-score characteristics. Therefore, we varied the rescaling factors along the horizontal and vertical axis of an image frame and noticed that the performance of the NOMF is nearly the same (difference of $<0.008$ of area under the curve) as the conventional filter (OMF) for $8\times6$ and $8\times3$ rescaling factors. It is worth mentioning that for other values of the rescaling factors, the area under the curve of the weighted F1-score for the images from both filters is lower than $0.23$.   

\subsubsection{Object Classification}
To assess the effect of approximation of NOMF on object classification, we have also classified vehicles into $4$ categories (Car, Bus, Track/Van and Bike) from the above traffic dataset following \cite{Hynna}. Objects in each frame are manually labelled and also the same objects across frames are assigned the same track identifier--this comprises the ground truth. We employ a modified LeNet5 architecture~\cite{Lecun} consisting of two convolutional ($5\times 5\times 6$ and $5\times 5\times 16$) layers and three fully connected (FC) layers ($120\times84\times4$) where the last layer had a softmax nonlinearity while others had ReLU. A $2\times2$ average pooling layer follows each convolutional layer. Since the LeNet5 architecture requires a fixed size input, we have selected a $42\times42$ image patch around the centre of a valid object  from EBBI. The $42\times42$ patches are filtered using both conventional median filter (OMF) and NOMF to produce two different datasets for the same LeNet5 model. For a fair comparison, the same LeNet5 model is trained and tested separately using $91885$ and $9063$ filtered images from the conventional median filter and NOMF, respectively, following the methodology used in \cite{Hynna}.

\begin{table}
\centering
\caption{Classification accuracy for test samples}
\begin{tabular}{ |M{2.5cm}||M{0.9cm}|M{0.9cm}|M{0.9cm}|M{0.9cm}|  }
 \hline
 Category& \multicolumn{2}{|c|}{per sample ($\%$)} &\multicolumn{2}{|c|}{per track ($\%$)}\\
 \hline
 & Median Filter &NOMF & Median Filter &NOMF\\
 \hline
 Car   &$83.72$    &$83.53$&$95.24$ &$95.24$\\
 Bus& $92.38$  & $86.72$   &$96.55$ &$96.55$\\
 Truck/Van &$67.39$ & $70.48$& $75$ &$78.12$\\
Bike    &$65.98$ & $63.07$&  $100$ &$100$\\
 \hline
% \textcolor{red}{Unbalanced accuracy} &\textcolor{red}{$82.31$} & \textcolor{red}{$79.88$}& \textcolor{red}{ $89.58$} &\textcolor{red}{$90.62$}\\
Balanced accuracy    &$75.25$ & $73.42$&  $90.52$ &$91.56$\\
 \hline
\end{tabular}
\label{tabaccuracy}
\vspace{-1.2em}
\end{table}
Table \ref{tabaccuracy} shows the per sample and per track accuracy for all test samples across different categories. To estimate per track accuracy, the final category is obtained by voting across the classification scores for all the images in a track across multiple frames\cite{Hynna}. We use balanced accuracy as a metric to overcome the imbalance in number of test samples across classes. Even though the balanced sample-wise classification accuracy using NOMF filtered images is $\approx 2\%$ lower than that using OMF, it can be seen that the track-wise accuracy for NOMF is slightly higher, as shown in the right column of Table~\ref{tabaccuracy}. While it could require more data to test if this difference is statistically significant, we can conclude that the system level performance of NOMF filtered images is similar to OMF ones.

\subsection{Energy Measurement}
As mentioned in Fig. \ref{fig:energy_denoise}, it is imperative to have a very low energy dissipation for the image denoise operation needed to detect blank image frames. For normal operation, the measured read and write energy of the memory, are $0.916$pJ and $6$pJ per bit at $1$ V respectively, close to the corresponding simulated values of $0.626$pJ and $5.1$pJ, respectively.

%\begin{figure}[t]
%\centering
%\includegraphics[scale=0.6]{image/CURRENT.eps}\\
%(a)\\
%\includegraphics[scale=0.6]{image/energyPixel.eps}\\
%(b)
%\caption{(a) Measured current (mA) consumption vs %frequency at different supply voltages. (b) %Corresponding energy consumption per pixel across %frequencies and supply voltages for $3\times 3$ filter %windows.}
%\label{CURRENT}
%\vspace{-1.2em}
%\end{figure}

\begin{figure}[b]
\centering
\includegraphics[scale=0.53]{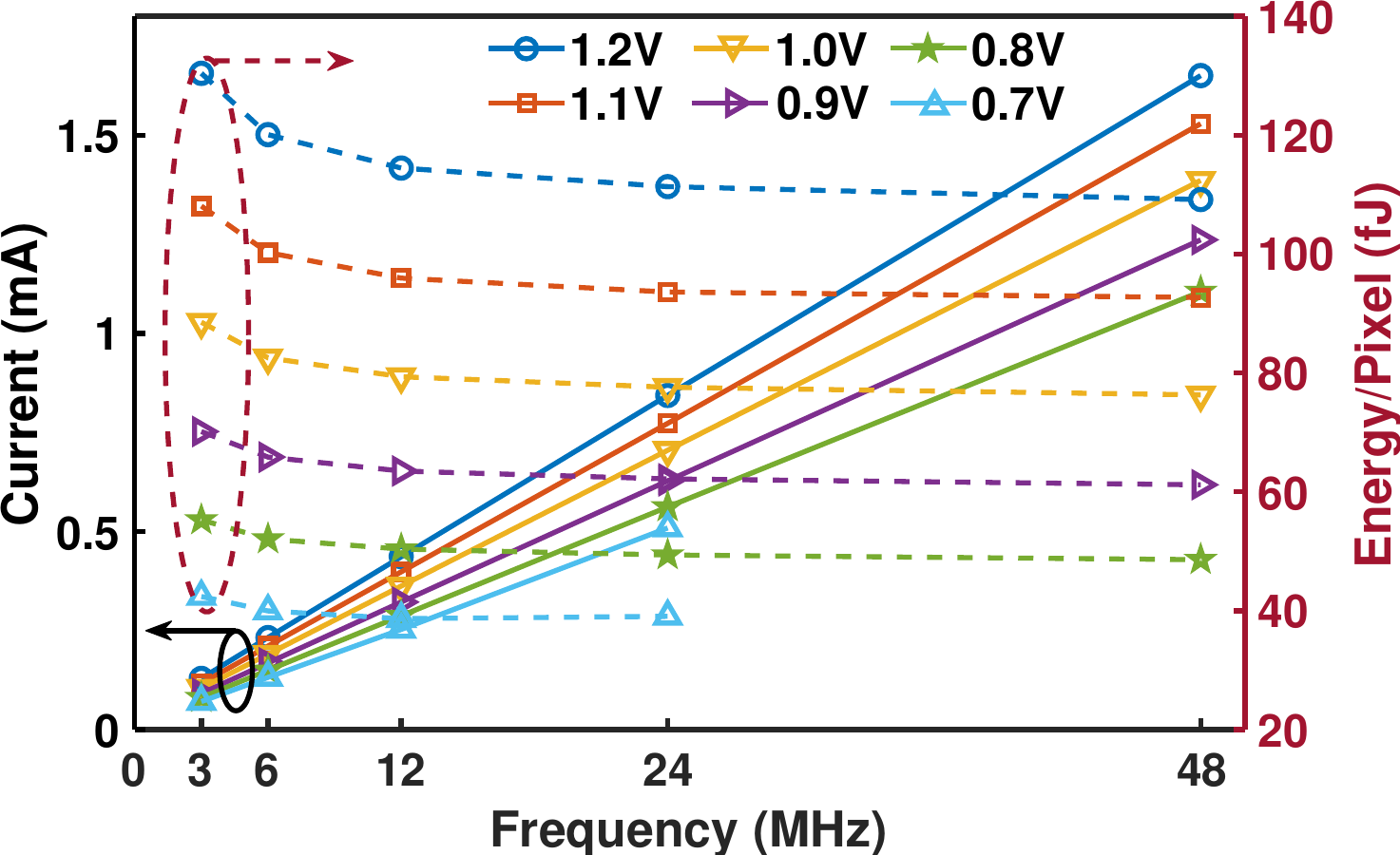}
\vspace{-0.5em}
\caption{Measured current (mA) consumption  (left Y axis) and energy/pixel (right Y axis) vs frequency at different supply voltages.}
\label{CURRENT}
\vspace{-1.3em}
\end{figure}

\begin{figure}[t]
\centering
\includegraphics[scale=0.33]{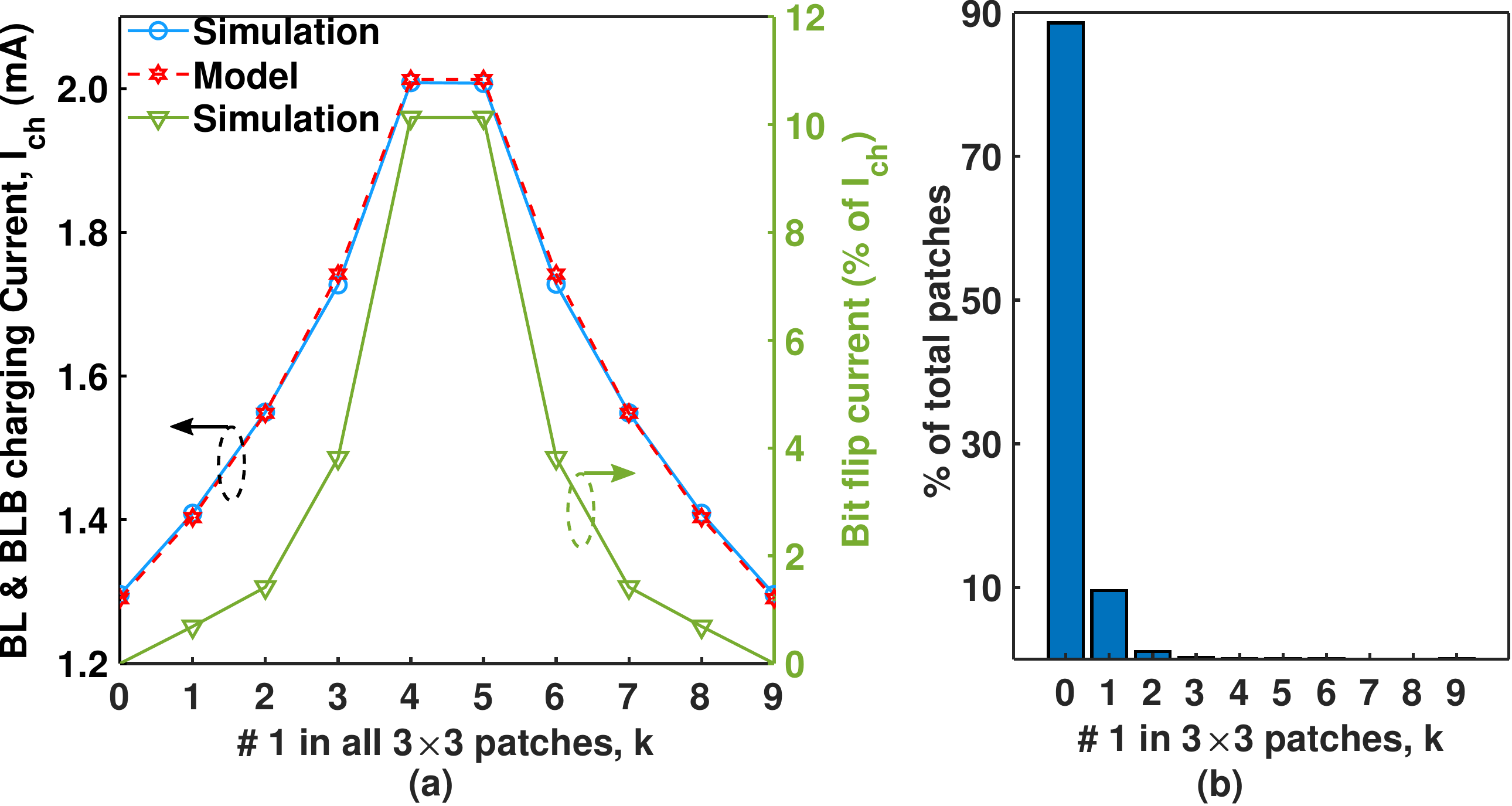}
\vspace{-0.5em}
\caption{(a) The left Y axis presents the simulated and modelled BL and BLB charging current, $\mathrm{I_{ch}}$ across different values of k at $VDD=1.2$V and $48$MHz, where k denotes the number of ``1"s in all $3\times3$ patches. The right Y axis presents the simulated bit flip current as a percentage of $\mathrm{I_{ch}}$ across different values of k. (b) Distribution of k: $88.5\%$ of $1.44$ billion ($0.3$M frames) $3\times3$ image patches are blank.}
\label{bl_blb_discharge}
\vspace{-1.2em}
\end{figure}

The more critical component is the energy dissipated for image filtering. Fig. \ref{CURRENT} (left Y axis) shows the average current consumption  across different frequencies  as well as supply voltages for this operation. It can be seen from the fig. that the IMF operates at power supply voltage downs to $0.7V$. However, the maximum operating frequency decreases due to read and write failure of configuration registers at this voltage. Note that, even though the current consumption is measured up to $48$MHz, the prototypes are operational at $70$MHz. The total current consumption of the test chip during NOMF operation can be formulated as:
\vspace{-0.3em}
\begin{align}
\vspace{-0.3em}
&I_{total}=I_{bitflip}+I_{leakage}+I_{IMF}+I_{ch}\notag\\
&\approx I_{IMF}
+((\rho+\lambda) N_{col} C_{BL(B)}+ nC_{WL})VDD\frac{f}{2}
\label{eqncurrent}
\vspace{-0.3em}
\end{align}
where $I_{bitflip}$ represents the current due to the bit flip operation during read-disturb, $I_{IMF}$ is the current consumed the digital controller and  $I_{ch}$ is the BL and BLB precharge current. Also, $N_{col}$ ($=320$) denotes the number of columns, $\rho$ ($=\frac{\Delta V_{BL}}{VDD}$) and $\lambda$ ($=\frac{\Delta V_{BLB}}{VDD}$) represent the fraction of BL and BLB discharge with reference to VDD, respectively. $I_{bitflip}$ can be neglected since its magnitude from simulations is small ($\approx 0.68\%$ of $I_{ch}$ at $k=1$ from Fig. \ref{bl_blb_discharge}(a)). $I_{leakage}$ denotes the leakage current, which is also negligible at higher operating frequencies ($<1\%$)--this is measured from the current consumed when no clock is supplied to the chip. 

To estimate the fraction of power dissipated in $I_{IMF}$, we use the above model in Eq.~(\ref{eqncurrent}) to estimate the last term $I_{ch}$ by calibrating the parameters from simulations and estimating the values of $\rho$ and $\lambda$ based on the input. For a blank (all ``0"s) or a full (all ``1"s) image patch, ($\rho,\lambda)=(1,0$) and ($\rho,\lambda)=(0,1$), respectively. However, in the case of a noisy image having $k$ pixels in each $n\times n$ window, we can show  that $1\leq \rho+\lambda \leq1+\beta\frac{ min(k,n^2-k)}{max(k,n^2-k)}$, where $\beta= 1-\frac{V_{trip}}{VDD}$ and $V_{trip}$ denotes the trip point voltage of a latch (back to back inverters) in the 6T-SRAM. It is seen from SPICE simulation  that $\beta=0.7$, $C_{WL}=330fF$ and $C_{BL}=C_{BLB}=140fF$. It is worth noting that the IMF controller takes two clock cycles to filter out $n$ rows for the ease of implementation. Hence, we use the factor $\frac{f}{2}$ in the formulation for total current. Implementation of denoising of $n$ rows in a single clock cycle would have doubled the throughput of the system. However, energy consumption would remain the same to a first order approximation. 
%Since most of the $n\times n$ patches of a image contain either zero active pixel or a noisy pixel, we can assume that $k\leq1$ (will find out from matlab simulation). 

The left Y axis of Fig.~\ref{bl_blb_discharge} (a) presents the simulated and modelled BL and BLB charging current for $n=3$ across different values of k at VDD=$1.2$V and $48$MHz, where k denotes the number of ``1"s in all $3\times3$ patches. We can see an excellent match across all values of $k$. Analyzing the image statistics from the dataset in~\cite{singla2020ebbinnot}, we can see from Fig.~\ref{bl_blb_discharge} (b) that only $9.5\%$ of total patches contain a single noisy pixel ($k=1$), resulting in $\overline{\rho+\lambda}\approx 1.01$. Hence, the average value of $I_{ch}$ can be estimated to be $\approx 1.331$ mA in real settings. Now, using this calibrated model from simulations, we can estimate from the measured results of Fig.~\ref{CURRENT} that the IMF controller current, $I_{IMF}$, contributes $\approx 36\%$ of the total current.

Fig. \ref{CURRENT} (right Y axis) presents the processing energy per pixel of the proposed prototype IC across different operating frequencies and supply voltages for the $3\times 3$ filter window. The figure illustrates that the leakage energy dominates at lower operating frequencies. However, at higher frequencies, the curves remain flat. It is worth mentioning that the proposed NOMF consumes $39$fJ to process a pixel at VDD=$0.7V$ for the $3\times 3$ filter window. Moreover, at higher filter sizes (e.g., $5\times5$), processing efficiency increases by $\approx \frac{5}{3}\times$. 
%\begin{figure}[t]
%\centering
%\includegraphics[scale=0.6]{image/energyPixel.eps}
%\caption{Measured energy consumption per pixel across frequencies and supply voltages for the $3\times 3$ filter window.}
%\label{energy}
%\end{figure}
\begin{table}[h]
\centering
\caption{Estimated and measured energy consumption of different filters}
\vspace{-0.5em}
%\resizebox{\columnwidth}{!}{
    \begin{tabular} {|M{2.2cm}|M{2.2cm}|M{2.2cm}|}\hline
          
Architecture & Energy(nJ) & \\ \hline
       MF &$191.67$ &Estimated\\ \hline
        MFRB &$117.72$ & Estimated\\ \hline
        IMC+NOMF &$1.68$ & Measured \\
        \hline
    \end{tabular}
\label{tabdata_energy}
\vspace{-1.2em}
\end{table}
\begin{figure}[b]
\centering
\includegraphics[scale=0.5]{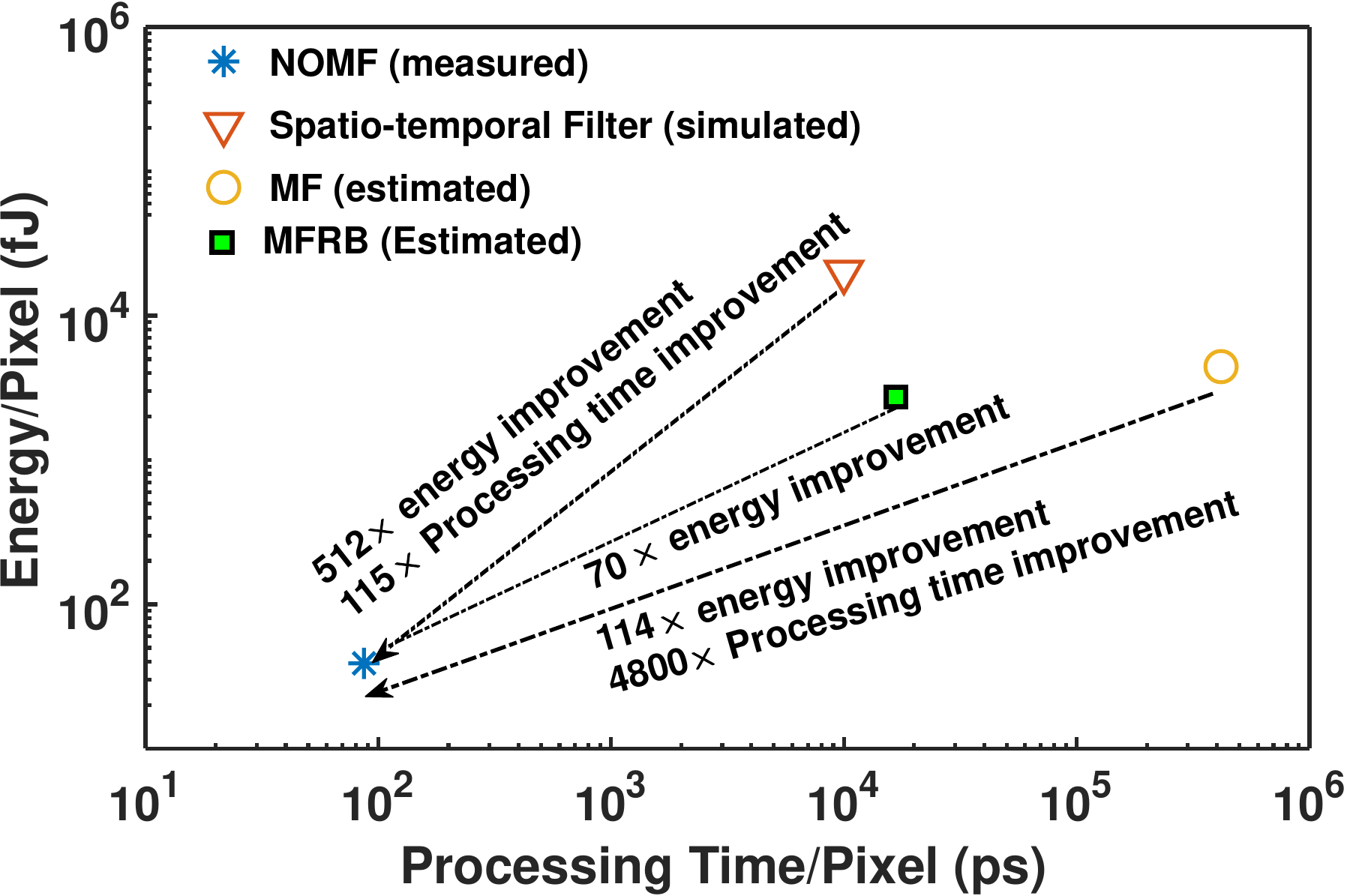}
\vspace{-0.5em}
\caption{Comparison of energy and processing time per pixel of NOMF with a spatio-temporal~\cite{temporal} filter and a digital implementation of the median filter.}
\label{comparison}
\vspace{-1.3em}
\end{figure}

\begin{table*}[h]
\centering
\caption{Comparison with different image filtering works}
\vspace{-0.5em}
%\resizebox{\columnwidth}{!}{
    \begin{tabular} {|M{3.5cm}|M{3.05cm}|M{3.05cm}|M{3.05cm}|M{3.05cm}|}\hline
          
 & This Work &~ISCAS'15\cite{temporal} &~TCAS-II'20\cite{9032196}  &ASP-DAC'20\cite{hashheat} \\ \hline
       Implementation \& Technology & ASIC, $65$nm & ASIC, $180$nm &ASIC, $65$nm &FPGA Artix-7\\ \hline
      Method & Spatial filter, in-memory computing & Spatio-temporal filter &Spatial filter, in-memory computing &Spatio-temporal filter\\ \hline
       
        Memory Size &$9.375$kB  &- &$9.375$kB & $0.25$kB\\ \hline
       Area/ cell ($\mu m^2$) &$3.6$  &$400$  &$3.54$ &- \\
        \hline
        Latency/ Pixel ($ns$) &$0.086$ at 24MHz &$10$  &$0.11$ &$10$ \\ \hline
         Power &$450 \mu W$ at 0.7V, 24MHz   &$1mW$ at 50Meps  &- &$0.471W$ \\
        \hline
    \end{tabular}
\label{compdenoising}
\vspace{-1.2em}
\end{table*}
\begin{table*}[t]
\centering
\caption{Comparison of different published IMC works}
\vspace{-0.5em}
 \begin{threeparttable}
%\resizebox{\columnwidth}{!}{
    \begin{tabular} {|M{3.6cm}|M{1.9cm}|M{1.9cm}|M{1.9cm}|M{1.9cm}|M{1.9cm}|M{1.9cm}|}\hline
          &This Work &ISSCC'19~\cite{twt} &TCAS-I'19~\cite{8787897} &~JSSC'18\cite{8226999} &SoVC'19~\cite{8778160} &JSSC'19~\cite{ABIS}\\ \hline
        Technology	 &$65$nm  &$55$nm  &$65$nm &$65$nm &$28$nm &$65$nm\\ \hline
        SRAM type &$6$T &Twin-$8$T &$6$T (split WL) &$6$T &$6$T (split-WL)  &$10$T \\ \hline
        Algorithm &Image denoising  &CNN &BNN &Versatile DNN &BNN  &BWN\\ \hline
         SRAM capacity &$9.375$kB &$0.468$kB &$2$kB &$102$kB &$2$kB &$2$kB\\ \hline
         Operating volatge (V) &$0.7$-$1.2$ &$1$ &$1$ &$0.55$-$1$ &$1$ &$0.8-1$\\ \hline
        input/weight precision &$1$\tnote{b}~/~$1$\tnote{b} &$1$-$4$~/~$2$-$5$ &$1$~/~$1$  &$1$~/~$1$ &$1$~/~$1$ &$6$~/~$1$\\ \hline
        Peak Throughput (GOPS) &$134.4$  &- &$278.2$ &$1380$ &$615$ &$8$\\\hline
       Peak Energy Efficiency (TOPS/W) &$51.3$  &$18.37$-$72.03$ [$13.15$-$51.5$]\tnote{a} &$30.49$-$55.8$  &$6.3$ &$300$ [$55.6$]\tnote{a} &$51.3$\\\hline
       Area Efficiency (TOPS/$mm^2$) &$0.243$  &- &$33.13$ &$0.089-0.365$ &$-$ &$0.127$\\\hline
    \end{tabular}
% }
\label{tabcomp}
\begin{tablenotes}
   \item[a] Assuming energy efficiency to $65$nm implementation following energy efficiency $\propto \frac{1}{(Tech. node)^2}$.
   \item[b] Assume the filter kernel is convoluted over the binary image and all the filter weights are $1$.
  \end{tablenotes}
  \end{threeparttable}
  \vspace{-1.8em}
\end{table*}

\subsection{Comparison}

Fig.~\ref{comparison} compares the performance of the proposed test chip (IMF) with a spatio-temporal filter~\cite{temporal} and digitally implemented median filter. Furthermore, several works on hardware implementation of image denoising are compared in Table~\ref{compdenoising}. Note that the spatio-temporal filter~\cite{temporal} works on the asynchronous events from an NVS, whereas the IMF and digital counterpart process event-based binary image. They reported $\approx 1$mW simulated power for $50$Meps (million events/second) resulting in $20$pJ/bit. Since the IMF (configuration register read and write operations) ceases to work beyond $24$MHz at $VDD=0.7V$, we assume the maximum operating frequency  to $24$MHz for the estimation of processing per pixel of the digital counterpart. We have estimated the energy and processing time for four different digital implementations of conventional median filter. (a) Median filter (MF): This architecture features single pixel read and process in every clock cycle-the execution time for this implementation is $(n^2+1)WH$ clock cycles ($W=$ image weight, $H=$ image height). This is a scalable architecture where more cycles are needed for bigger images but the hardware cost is fixed. (b) Median filter with pipelined read (MFPR): This architecture has a pipelined design and reads W pixels simultaneously. This implantation takes $2nH$ clock cycles to denoise an image and requires W processing elements (adder). It is to be noted that for different values of n, we need different n-bit adders. This results in a higher silicon area; it is also difficult to scale up with a higher resolution image due to area constraints. (c) Median filter with Row Buffering (MFRB): This design reads a single pixel at a time and has n-1 row buffers of width $W$. In this case, $n-1$ consecutive rows of an image are stored in the buffer and used in the subsequent filtering operation while stride moves in the vertical direction. The oldest row is replaced by the new row read. This will save 2n memory read energy. (d) Median filter with pipelined read and row buffer (MFPRRB): The last architecture combines both pipelining and row buffering. This  parallel-pipelined and row buffered implementation takes approximately $2H$ clock cycles to process an image. Since IMF takes two clock cycles to process $n$ rows, the proposed design takes $\frac{2H}{n}$ cycles to denoise an image. Therefore the IMF is $3-4800\times$ faster than the digital implementation of the median filter for $n=3$.

We have estimated the energy/bit of the convention median filter from the read ($0.916$pJ) and write energy ($6$pJ) measurements of the designed $320\times240$ SRAM memory. The increased size $6$T-SRAM designed for NOMF has $1.57\times$ (140fF/89fF) higher parasitic capacitance on BL lines. Therefore, during the energy estimation of the convention median filter, we have taken into account the lower parasitic capacitance. For a fair comparison, we scale down the estimated energy consumption of the digital implementation at $0.7V$. Table~\ref{tabdata_energy} compares the energy consumption of different filter implementations. Please note that we only consider the memory read and write energy of a single-pixel of the designed chips to estimate the conventional median energy. Even though the memory read energy will be lower by $n$ for the parallel read implementations (MFPR, MFPRRB), and the energy dissipated by the processing element will nullify the benefit to some extent.

It can be seen from Fig.~\ref{comparison} that the IMF achieves $512\times$ and $>70\times$ energy improvement over the state-of-the-art spatio-temporal filter and digital implementation, respectively. 

Similarly, it can be estimated from Fig.~\ref{fig:energy_denoise} that the NOMF achieves $1.24\times$ energy improvement compared to the conventional median filer in object recognition system. Considering the per frame execution time of~\cite{nverma} at $70$MHz, the object recognition system along with NOMF is $1-3.74\times$ faster than that with the conventional median filter. Therefore, NOMF achieves $\sim1.24-4.6\times$ energy-delay-product improvement at the system level.

Table \ref{tabcomp} compares the performance of IMF with the recently published IMC works on Binary Weighted Network (BWN: stored weights in the SRAM are binary) and Binary Neural network (BNN: inputs and stored weights are binary). A major difference of this work with other neural networks is that we store the image in the memory and operate on it, while for the neural networks, the weights are typically stored in memory and the image pixels are streamed in as inputs. To calculate the peak throughput and energy efficiency, we assume the filter kernel is convoluted over the binary image, and all the filter weights are ``1". Since the IMF processes $320\times3$ pixels in a cycle, the peak throughput is $2\times320\times3\times70$M$~=134.4$ GOPS at $70$ MHz. Furthermore, minimum energy consumption, $39$fJ per pixel, translates to $51.3$ TOPS/W energy efficiency at $VDD=0.7V$ (2 operations/pixel), which is comparable with the state-of-the-art. Note that the peak throughput depends on the operating frequency, image patch size, level of parallelism. Hence, it can be enhanced further by increasing the patch size at the cost of more approximation or by separating the memory into multiple sections along the column to increase the level of parallelism. Moreover, the memory segmentation and parallelism also amortize the same BL charging energy into the computation of multiple patches , which will further enhance energy efficiency.

\section{Discussion}
\label{validframedetection}
\subsection{Valid frame Detection}
%\subsection{Valid frame Detection}
Fig. \ref{fig:energy_denoise} demonstrated the benefits of leveraging temporal sparsity in video streams to reduce system level energy dissipation. The key processing block to achieve this when using NVS is the noise filter. However, after filtering, a circuit is still needed to detect that the frame is blank, i.e. all pixel values are `0'. To retain the energy efficiency of the system, this computation should also be performed in or near memory. While we do not have this function integrated in our chip, it can be easily done by tracking the bitline values during the filtering operation.

\begin{figure}[t]
\centering
\includegraphics[scale=0.38]{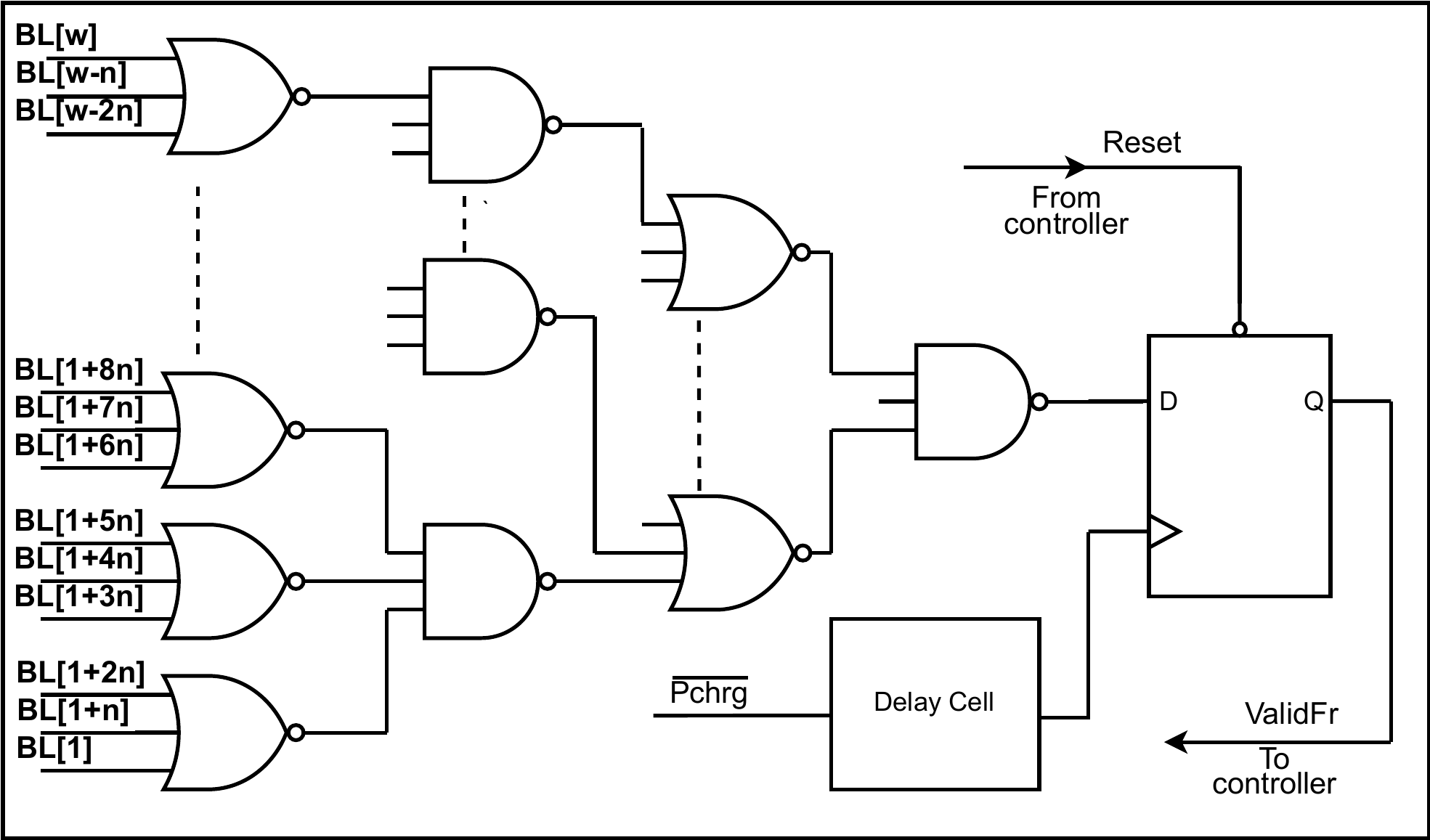}
\vspace{-0.5em}
\caption{Circuit diagram of a valid frame detector.}
\label{zerodetect}
\vspace{-1.3em}
\end{figure}

Fig.~\ref{zerodetect} presents a circuit diagram of a valid frame detector consisting of three input  $\ceil*{\frac{w}{3n}}+\ceil*{\frac{w}{27n}}+\cdots$ NOR gates, $\ceil*{\frac{w}{9n}}+\ceil*{\frac{w}{81n}}+\cdots$  NAND gates, and one DFF and delay cell for sensing $w$ BLs, where $w$ denotes the width of an image. Since BLs do not get discharged at the presence of an object, they can be ORed together to detect a valid frame when the image filtering is being done row by row. %Besides, $n$ BLs are connected for an $n\times n$ filter window leading to a reduction of the number of BLs to be sensed by a factor of $n$.
At the beginning of the filtering operation, the DFF is reset by the Filter signal. Once the ValidFr signal goes high, the IMF controller can register it as a presence of an object. The clock signal of the DFF is a delayed and inverted version of the Pchrg signal (Filter and Pchrg signals are shown in  Fig.~\ref{fig:timingdiagram}) to take into account the settling time of BL and BLB lines. The valid frame detector consumes $3.3$pJ ($0.2\%$ of total energy consumption) for an entire frame in simulation, and the area overhead is $335\mu m^2$ ($0.06\%$ of total area) for $w=240$ and $n=3$. %The parasitic capacitance on a BL due to a NOR gate is $0.5fF/n$, which is negligible compared to $C_{BL}\approx140$fF.%the extra capacitance on BLBs (discussed in section~\ref{imagede}) to some extent.  %Since it is discussed in section \label{imagede} that the BLB experiences $\approx4.08$fF higher capacitance than the BL, sensing BLs nullifies the extra capacitance on BLBs to some extent.

\subsection{Majority Function}
The operation done by this SRAM macro can be generalized to the majority function. Hence, apart from NOMF, it can also be used to implement adder circuits~\cite{NAVI2009126} which are widely used for digital arithmetic operations.

In general, while this macro does not perform other programmable filter kernels, it is intended to replace the SRAM used to store the input image and do in-place filtering of the same. Any image processing hardware would require SRAM to store the image and we propose to make that SRAM functional instead of just storing the data.

It is worth mentioning that the proposed architecture does not support multi-bit images. However, the RGB images can be transformed to event-based binary images by detecting changes (events) at the same pixel across frames comparing both linear and logarithmic pixel intensities~\cite{9162948} and apply the proposed method for denoising to get the energy savings in the subsequent  image processing steps for multi-bit operation. 
%Operation mode &Current summation & Current summation & Current summation &Digital XNOR &Current summation & Charge distribution\\ \hline
\section{Conclusion}
\label{conclu}
This paper presents a $6$T-SRAM based image denoising hardware architecture for event-based binary image that exploits read-disturb operation. The proposed hardware consumes as low as $39$fJ to process a pixel and takes $1.7\mu$s to denoise a $240\times180$ image frame. We demonstrated that the test chip achieves a  $>210\times$ improvement of energy-delay product (EDP) compared to a fully digital implementation of the conventional median filter. This enormous gain of EDP can be attributed to the NOMF algorithm and in-memory computing, which enables the parallelism in denoising of an image. Even though NOMF introduces approximation in computation, we show that the system level metrics of object detection and tracking are minimally affected compared to the conventional median filter. The extremely low energy dissipation indicates that the proposed in-memory computing based image denoising hardware can be the right candidate for triggering ON more energy hungry object detection CNNs while maintaining low energy for blank frames and thus exploiting the temporal sparsity of video streams in IoT.

%\appendix[$7$-T implementation]
% or


%\appendices

%\section{}

%\section*{Acknowledgment}
\ifCLASSOPTIONcaptionsoff
  \newpage
\fi



%\bibliographystyle{IEEEtran}
%begin{thebibliography}{1}


%\end{thebibliography}
\bibliographystyle{ieeetr}
\footnotesize
\bibliography{irpn}

\end{document}